\documentclass[article,12pt]{article}

\usepackage[a4paper, top=1in, bottom=1in, left=1in, right=1in]{geometry}

\usepackage{siunitx}
\usepackage[utf8]{inputenc}

\usepackage{mathtools} 
\usepackage{amsmath} 
\usepackage{dcolumn}
\usepackage{revsymb}
\usepackage{bm}
\usepackage{bbm}
\usepackage[left]{lineno}
\usepackage{setspace,caption}
\usepackage{authblk}
\usepackage{graphicx}
\usepackage{helvet}
 
\usepackage[helvet]{sfmath}
\usepackage[T1]{fontenc}
\usepackage{xurl}
\DeclareUnicodeCharacter{03B1}{$\alpha$}
\DeclareUnicodeCharacter{03BC}{$\mu$}

\def\Be{Be$^{+}$}
\def\Arforty{${}^{40}$Ar$^{13+}$}
\def\Arthirtysix{${}^{36}$Ar$^{13+}$}
\def\Ar{Ar$^{13+}$}
\def\Yb{${}^{171}$Yb$^{+}$}

\def\shalf{${}^{2}$S$_{1/2}$}
\def\phalf{${}^{2}$P$_{1/2}$}
\def\pthreehalves{${}^{2}$P$_{3/2}$}

\def\fsevenhalves{${}^{2}$F$_{7/2}$}

\newcommand{\balpha}{\bm{\alpha}}

\newcommand{\bfr}{{\bm {r}}}

\newcommand{\bfp}{{\bm {p}}}

\newcolumntype{w}[1]{D{.}{.}{#1}}

\newcolumntype{.}{D{x}{}{-1}}

\let\oldequation\equation
\let\oldendequation\endequation
\renewenvironment{equation}
  {\linenomathNonumbers\oldequation}
  {\oldendequation\endlinenomath}

\begin{document}

\title{An Optical Atomic Clock Based on a \par Highly Charged Ion}

\author[1,5,$\dagger$]{\mbox{Steven A. King}}

\author[1,$\dagger$,*]{\mbox{Lukas J. Spieß}}

\author[1,2,6]{\mbox{Peter Micke}}

\author[1]{\mbox{Alexander Wilzewski}}

\author[1,7]{\mbox{Tobias Leopold}}

\author[1]{\mbox{Erik Benkler}}

\author[1]{\mbox{Richard Lange}}

\author[1]{\mbox{Nils Huntemann}}

\author[1,3]{\mbox{Andrey Surzhykov}}

\author[1]{\mbox{Vladimir A. Yerokhin}}

\author[2]{\mbox{Jos\'e R. Crespo L\'opez-Urrutia}}

\author[1,4,*]{\mbox{Piet O. Schmidt}}

\affil[1]{Physikalisch-Technische Bundesanstalt, Braunschweig, Germany}

\affil[2]{Max-Planck-Institut f\"ur Kernphysik, Heidelberg, Germany}

\affil[3]{Technische Universit\"at Braunschweig, Braunschweig, Germany}

\affil[4]{Institut f\"ur Quantenoptik, Hannover, Germany}

\affil[5]{Present Address: Oxford Ionics, Begbroke, United Kingdom}
\affil[6]{Present Address: CERN, Geneva, Switzerland}
\affil[7]{\vspace{2ex} Present Address: LPKF Laser \& Electronics AG, Garbsen, Germany}

\affil[*]{Corresponding authors. Email: lukas.spiess@quantummetrology.de, piet.schmidt@quantummetrology.de}

\affil[$\dagger$]{These authors contributed equally to this work.}

\date{\vspace{-5ex}}

\maketitle

\textbf{
Optical atomic clocks are the most accurate measurement devices ever constructed and have found many applications in fundamental science and technology \cite{ludlow_optical_2015, safronova_search_2018, mehlstaubler_atomic_2018}. The use of highly charged ions (HCI) as a new class of references for highest accuracy clocks and precision tests of fundamental physics \cite{kozlov_hci, schiller_hydrogenlike_2007, berengut_enhanced_2010, berengut_optical_2012, derevianko_highly_2012, safronova_highly_2014-1, yudin_magnetic-dipole_2014, beloy_quadruply_2020} has long been motivated by their extreme atomic properties and reduced sensitivity to perturbations from external electric and magnetic fields compared to singly charged ions or neutral atoms. Here we present the realisation of this new class of clocks, based on an optical magnetic-dipole transition in \Ar{}. Its comprehensively evaluated systematic frequency uncertainty of $2.2\times 10^{-17}$ is comparable to that of many optical clocks in operation. From clock comparisons we improve by eight and nine orders of magnitude upon the uncertainties for the absolute transition frequency \cite{egl_application_2019} and isotope shift ($^{40}$Ar vs.~$^{36}$Ar)\cite{orts_exploring_2006}, respectively. These measurements allow us to probe the largely unexplored quantum electrodynamic nuclear recoil, presented as part of improved calculations of the isotope shift which reduce the uncertainty of previous theory \cite{zubova_isotope_2016} by a factor of three. This work establishes forbidden optical transitions in HCI as references for cutting-edge optical clocks and future high-sensitivity searches for physics beyond the standard model.}

\section{Introduction}
\label{sec:intro}
In highly charged ions (HCIs), the outer electron is strongly bound and thus experiences extremely magnified quantum electrodynamic (QED) and nuclear size effects \cite{gillaspy_highly_2001}. %
Binding energy, fine-structure and hyperfine splitting steeply scale up with the total charge of the ion. All this makes HCI extremely sensitive probes for testing fundamental physical theories \cite{safronova_search_2018, kozlov_hci}. Many of the atomic properties of HCI offer advantages for clock applications \cite{kozlov_hci}. High binding energies shift electric-dipole-allowed ($E1$) transitions out of the optical range. This suppresses the sensitivity of the outer electrons to external electromagnetic perturbations \cite{schiller_hydrogenlike_2007, derevianko_highly_2012, yudin_magnetic-dipole_2014} by orders of magnitude in comparison to even the most insensitive species currently used in optical atomic clocks \cite{arnold_blackbody_2018, hachisu_trapping_2008}. \par

For the proof-of-principle demonstration of an HCI optical clock, we chose boron-like (B-like) \Ar{}. With five bound electrons, its ground state ($1s^{2}\,2s^{2}\,2p_{1/2}$) is linked to the lowest excited state by a \phalf{} $\rightarrow$ \pthreehalves{} magnetic-dipole ($M1$) fine-structure transition with a wavelength of 441~nm. This transition is well-suited for an initial demonstration as it is the most precisely measured transition in any HCI.
For much  of the last two decades, the most precise measurements of optical transitions in HCI relied on the electron beam ion trap (EBIT) for HCI production and trapping. The use of grating spectrometers \cite{bieber_studies_1997, orts_exploring_2006, soria_orts_zeeman_2007}, and laser spectroscopy \cite{mackel_laser_2011} both afforded a Doppler-limited fractional uncertainty of $3 \times 10^{-7}$ for \Ar{}, limited by the high temperature ($T>10^5$\,K) of HCI trapped in EBITs. A recent experiment improved this by applying negative electronic feedback in a Penning trap for cooling \Ar{} down to the kelvin-level. Usage of the continuous Stern-Gerlach effect for detection of the HCI's internal state after laser excitation yielded a fractional uncertainty of $9 \times 10^{-9}$ \cite{egl_application_2019}. Still, an uncertainty gap of ten orders of magnitude persisted in comparison to state-of-the-art optical clocks \cite{brewer_27+_2019}.\par

Recently, novel techniques have overcome several key hindrances for HCI-based clocks: sympathetic cooling of hot HCI from megakelvin down to millikelvin temperatures using Be$^{+}$ ions in a Paul trap \cite{schmoger_coulomb_2015}; readout of the HCI internal state by quantum logic spectroscopy (QLS) \cite{micke_coherent_2020}, and full quantum control over the motional state of a trapped HCI \cite{king_algorithmic_2021}. \par

These breakthroughs, combined with the techniques and characterisation presented here, have now enabled us to realise an HCI-based clock for the first time.
Comprehensive measurements and analysis of all the systematic shifts of the clock transition result in a total fractional systematic uncertainty of $2.2\times10^{-17}$, with most shifts being below the $10^{-18}$ level. We determined the absolute frequency of the transition with a total fractional uncertainty of $1.5\times10^{-16}$ by comparing it against a local optical clock using a single \Yb{} ion and by using a previous measurement of its absolute frequency \cite{lange_improved_2021}. Measurements performed using \Arforty{} and \Arthirtysix{} yielded the isotope shift, with an uncertainty that reveals QED nuclear recoil effects for the first time in a many-electron system. We therefore present improved calculations of the shift which reduce the uncertainty by a factor of three over previous work \cite{zubova_isotope_2016}, showing excellent agreement with the experimental value. Moreover, we extract the Land\'e $g$-factor of the \pthreehalves{} state with higher accuracy than in earlier work \cite{micke_coherent_2020}, and establish an upper bound for the quadrupole moment of \Ar{} that confirms reported calculations \cite{agababaev_ground-state_2018,agababaev_g_2019, yu_investigating_2019, robert_priv_2021}. We also identify in this proof-of-principle experiment that an improved ion trap will bring the total systematic uncertainty below the $10^{-18}$ level. 

\section{Experimental setup}
\label{sec:experiment}

HCI are produced in a miniature EBIT \cite{micke_heidelberg_2018}, and transferred through an electrodynamic beamline to a linear Paul trap held at a temperature near 4\,K \cite{leopold_cryogenic_2019, micke_closed-cycle_2019}. There, a two-ion crystal composed of a single \Ar{} together with a single $^{9}$Be$^{+}$ ion is prepared. The latter is used for sympathetic cooling to below the millikelvin level and determination of the internal state of the HCI by QLS \cite{schmidt_spectroscopy_2005, micke_coherent_2020}. In the trap we reach a residual gas pressure below $10^{-12}$~Pa, suppressing charge-exchange collisions and enabling a mean HCI storage time of around 100~minutes.\par

\begin{figure*}[t]
	\includegraphics[width=\textwidth]{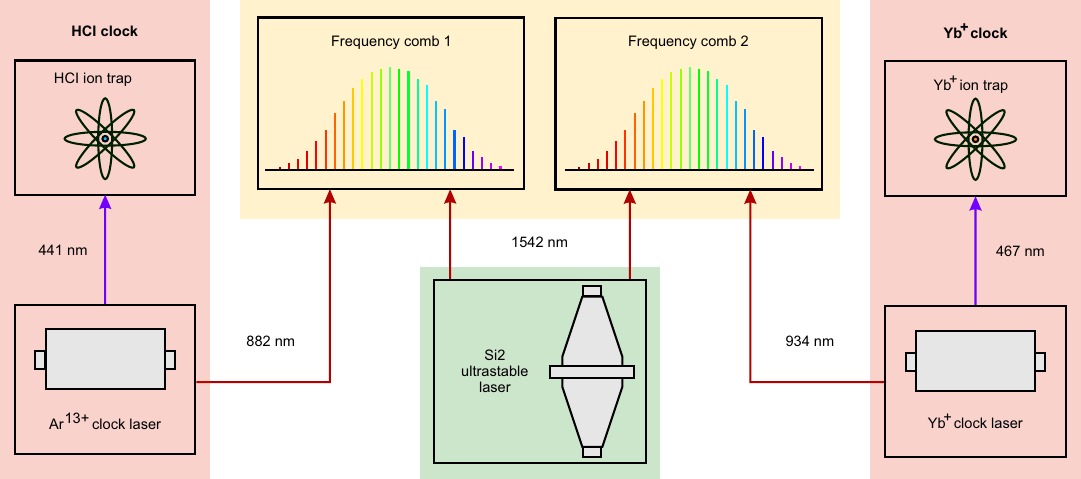}
	\caption{\textbf{Scheme of the method for optical frequency comparison.} Each of the two clock lasers (\Ar{}, \Yb{}) is locked for pre-stabilisation to its own local cavity and frequency comb, and ultimately steered to the corresponding optical transition by a digital control loop. The two frequency combs are locked to the exceptionally stable cryogenic silicon cavity Si2 \cite{matei_1.5_2017}. This method yields for each comb the frequency ratio between its clock laser and the Si2-stabilised laser. The dedicated laboratories are linked through phase-stabilised optical fibres \cite{benkler_end--end_2019}.}
	\label{fig:simplifiedclocklaser}
\end{figure*}

The light for interrogation of the \Ar{} clock transition is based on an external cavity diode laser with a wavelength of 882~nm. We pre-stabilise it with a local reference cavity and further by an optical frequency comb (see Fig.~\ref{fig:simplifiedclocklaser}) via a frequency-steering acousto-optical modulator (AOM). The frequency comb is locked via a \SI{1.5}{\micro\meter} reference laser to the cryogenic silicon cavity Si2 \cite{matei_1.5_2017}, resulting in a fractional frequency stability of $4 \times 10^{-17}$ at timescales of $1\dots 100$\,s. The clock output is generated by adjusting the offset between frequency comb and the pre-stabilised laser. Frequency-doubling of the 882~nm light then produces the 441~nm light for driving the clock transition (for details, see the Methods section).\par

The clock cycle begins with a 38~ms-long sequence of laser pulses that performs the following steps. First, the ion crystal is cooled close to the motional ground state of the two axial normal modes using resolved-sideband cooling on the \Be{} ion. The internal state of the HCI is then prepared using quantum-logic assisted optical pumping \cite{micke_coherent_2020}, followed by algorithmic cooling of the two radial normal modes in which the HCI has a significant amplitude of motion \cite{king_algorithmic_2021}.\par
After this preparation sequence, the clock transition is interrogated using a 15~ms-long pulse from the 441~nm laser, which leads to an interaction-time-limited linewidth of approximately 50~Hz. This interrogation time is approximately 1.5 times the natural lifetime of the excited electronic state \cite{lapierre_relativistic_2005}, and leads to the lowest achievable statistical uncertainty for our experimental parameters \cite{peik_laser_2006}.\par

We employ QLS to map the internal state information of the HCI after interrogation onto the logic ion Be$^+$, where it is detected using the fast cycling transition \cite{schmidt_spectroscopy_2005}. Using the result of repeated interrogations, the offset between the clock output and the line centre of any particular Zeeman component of the clock transition is steered with a dedicated second-order integrating digital control loop (in the following dubbed `servo') \cite{peik_laser_2006}.\par

Gradual charging of surfaces close to the trapped HCI by the ultraviolet laser beams used for the cooling and manipulation of the \Be{} ion can induce slow position drifts of the ion, leading to a first-order Doppler shift \cite{rosenband_frequency_2008}. We suppress it by averaging the servo outputs for two counter-propagating clock beams for each Zeeman component. Consequently, eight independent servos run in parallel, two for each of the four Zeeman components displayed in Fig.~\ref{fig:transitions}. As described in the Supplementary Material, we suppress additional small systematic drifts by randomising the order in which the interrogations for the eight servos are performed. The average of the eight servo outputs is free of the linear Zeeman shift, first-order Doppler shift, and quadrupole shift \cite{dube_evaluation_2013} and steers the clock output. Its frequency is then directly compared to that of another clock by means of the optical frequency comb setup shown in Fig.~\ref{fig:simplifiedclocklaser}, with only minor corrections necessary for those systematic shifts that are not averaged to zero. \par

\begin{figure}[t]
   \begin{center}
   	\includegraphics[width=0.9\textwidth]{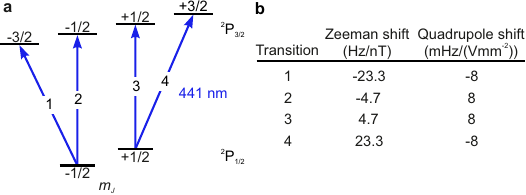}
    	\caption{\textbf{Zeeman components used in this work for the \Ar{} clock.} \textbf{a}, simplified term scheme of the transitions with magnetic quantum numbers ($m_J$) of their ground and excited states. \textbf{b}, Corresponding magnetic sensitivities (Zeeman shift) and maximum sensitivities to electric field gradients (quadrupole shift) \cite{yu_investigating_2019}. During operation, a static magnetic field of \SI{23}{\micro\tesla} is applied for defining a quantisation axis. The high charge state of the \Ar{} ion allows our trap to work with a gradient of approximately 2~Vmm$^{-2}$, much lower than those needed in other linear ion traps used for optical clocks.}
    	\label{fig:transitions}
    \end{center}
\end{figure}

In addition to the dominant $^{40}$Ar isotope, we are also able to run the clock with the rare $^{36}$Ar isotope. Details on its preparation are given in the Methods section.\par

\section{Frequency measurement}
\label{sec:freq}

\Ar{} features a comparatively broad natural linewidth of $17\,$ Hz for the clock transition, leading to higher frequency instability compared to other optical clocks. Nevertheless, it benefits from many of the advantageous properties of HCIs such as a small differential polarisability and quadrupole moment. The dominant systematic shift in our system is caused by residual driven motion at the frequency of the ion trap's radially confining potential, referred to as excess micromotion, which leads to time-dilation relative to the laboratory frame. In the present case, an unexpected component along the axial direction of the trap that cannot be compensated was probably caused by a misalignment of the trap electrodes \cite{leopold_cryogenic_2019}. Here it is by far the largest frequency shift, and limits the relative uncertainty to $2.2\times10^{-17}$. However, this shift is not exacerbated by the large charge state of the HCI since it depends only on the secular frequency of the ion in the confining potential, which we chose to be similar to that used for singly charged ions.
An improved ion trap will reduce this uncertainty to well below the $10^{-18}$ level \cite{keller_precise_2015}. The next largest systematic uncertainty is at the low $10^{-18}$ level and arises from the a.c. Zeeman shift induced by the clock laser off-resonantly coupling to neighboring Zeeman states. The uncertainties on all other investigated systematic shifts, including time dilation shifts from residual thermal motion, shifts from the electric quadrupole moment of the excited \pthreehalves{} clock state coupling to an electric field gradient, and higher order magnetic field shifts, are even smaller under our operating conditions.\par

As reference for the frequency of the \Ar{} clock, an optical clock based on the \shalf{} $\rightarrow$ \fsevenhalves{} electric octupole ($E3$) transition in \Yb{} is employed. This system has a fractional systematic uncertainty of $3\times10^{-18}$ and a fractional frequency instability of $1\times 10^{-15}(\tau/\mathrm{s})^{-1/2}$, where $\tau$ is the averaging time in seconds. Since the \Yb{} and \Ar{} ions are trapped at unequal heights, a correction for the  difference in gravitational potential is applied.\par

The outputs of the \Yb{} and \Ar{} clocks were delivered to separate optical frequency combs through phase-stabilised optical fibre links, as shown in Fig.~\ref{fig:simplifiedclocklaser}. Each comb allows a direct measurement of the local ratio between the frequencies of the clock and the ultrastable \SI{1.5}{\micro\meter} laser \cite{matei_1.5_2017}. Combination of the synchronous measurements of the two ratios yielded the ratio between the two clock laser frequencies.\par

Table~\ref{tab:systematics} summarises the dominant systematic frequency shifts on the \Ar{} clock transition and their uncertainties as well as those arising from the measurement of the absolute transition frequency. An in-depth analysis of all of the systematic shifts pertinent to the presented measurements can be found in the Supplementary Material.

\begin{table}
	\begin{center}
	    \caption{\textbf{Systematic shifts ($\Delta\nu$) and corresponding 1-$\sigma$ uncertainties ($\sigma$)}}
		\begin{tabular*}{\textwidth}{c @{\extracolsep{\fill}} cccc}
		    \hline
			 & \multicolumn{2}{c}{$\Delta\nu/\nu$ ($10^{-17}$)}  & \multicolumn{2}{c}{$\sigma/\nu$ ($10^{-17}$)}\\
			 Shift & \Arforty{} & \Arthirtysix{} & \Arforty{} & \Arthirtysix{} \\
			\hline
			Excess micromotion & $-44.3$ & $-55.0$ & $2.2$ & $2.1$ \\
			Laser-induced a.c. Zeeman & \multicolumn{2}{c}{$0$} & \multicolumn{2}{c}{$0.2$}\\
			Secular motion & \multicolumn{2}{c}{$-0.1$} & \multicolumn{2}{c}{$<0.1$}\\
			Quadrupole shift & \multicolumn{2}{c}{$0$} & \multicolumn{2}{c}{$<0.1$} \\
			Quadratic Zeeman & \multicolumn{2}{c}{$<0.1$} & \multicolumn{2}{c}{$\ll0.1$}\\
			\hline
			\textbf{Total \Ar clock} & $-44.4$ & $-55.1$ & $2.2$ & $2.1$\\
		    \hline
			Statistics & \multicolumn{2}{c}{$0$} & $10$ & $13$\\
			Uncompensated paths & \multicolumn{2}{c}{$0$} & \multicolumn{2}{c}{$0.1$}\\
			Yb$^{+}$ $E3$ absolute frequency & \multicolumn{2}{c}{$0$} & \multicolumn{2}{c}{$13$}\\
			Gravitational redshift & \multicolumn{2}{c}{$16.4$} & \multicolumn{2}{c}{$0.6$}\\
			\hline
			\textbf{Total absolute frequency} & $-28$ & $-39$ & $15$ & $17$\\
			\hline
		\end{tabular*}
		\caption*{Values are given relative to the transition frequency $\nu$. When applicable, differing values for \Arforty{} and \Arthirtysix{} are given. The different values for the excess micromotion shift are primarily caused by the difference in isotopic mass. The absolute frequencies are limited by the statistical uncertainties and the total uncertainty of the previous measurement of the absolute frequency of the \Yb{} clock transition, which contribute approximately equally. Further systematic shifts are at the $10^{-19}$ level or below (see Extended Data Table 2 and the Supplementary Material).}
		\label{tab:systematics}
	\end{center}
\end{table}

\section{Results and analysis}

\begin{figure}[t]
	\includegraphics[width=0.9\textwidth]{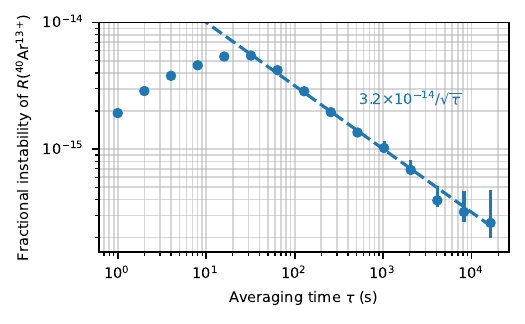}
	\caption{\textbf{Instability of the ratio between the \Arforty{} and \Yb{} transition frequencies}. Statistical uncertainty of the ratio $R$(\Arforty{}) $=\nu$(\Arforty{}) / $\nu$(\Yb{} $E3$) as a function of averaging time $\tau$, as measured using the overlapping Allan deviation. The error bars correspond to the 68\% confidence interval. The instability improves as $\tau^{-1/2}$ beyond the servo time constant of around 30 seconds, as characteristic for white frequency noise. The dashed line is a fit to the data for averaging times beyond 300~s.}
	\label{fig:ratioadev}
\end{figure}

The ratio between the \Ar{} and \Yb{} $E3$ transition frequencies were repeatedly measured over several days for each of the two Ar isotopes, with a total measurement time of approximately 100,000~s for \Arforty{} and 50,000~s for \Arthirtysix{}. The frequency instability of the complete dataset for \Arforty{} is shown in Fig.~\ref{fig:ratioadev}. At timescales beyond the servo time constant of 30~s, we infer a frequency instability of $3.2\times10^{-14}\,(\tau/\mathrm{s})^{-1/2}$. For \Arthirtysix{}, a lower instability of $2.6\times10^{-14}\,(\tau/\mathrm{s})^{-1/2}$ was obtained due to more efficient state preparation of the ions, lowering the dead time of the experimental cycle. We do not observe a floor in the frequency instability out to averaging times of more than 10,000~s, as expected for white frequency noise. Extrapolating the observed instability to the total measurement time resulted in a fractional statistical uncertainty of $1 \times 10^{-16}$ for each of the two frequency ratios.

The final results are summarised in Extended Data Table 1. The absolute values of the \Arforty{} and \Arthirtysix{} clock-transition frequencies are derived from the absolute frequency of the \Yb{} $E3$ transition reported in Ref.~\cite{lange_improved_2021}. The overall fractional frequency uncertainties of $1.5$ and $1.7\times10^{-16}$, respectively, place these results amongst the most accurately reported absolute transition frequencies of all time \cite{beloy_frequency_2021,nemitz_absolute_2021,lange_improved_2021,pizzocaro_absolute_2020}. For \Arforty{} the value excellently agrees with the most recently published measurement \cite{egl_application_2019}, while surpassing its accuracy by almost eight orders of magnitude.\par

Combining the data for the two isotopes allowed us to determine the \Arforty{}\,--\,\Arthirtysix{} isotope shift with a fractional uncertainty of $5.7\times10^{-11}$, where some systematic uncertainties were common-mode rejected. Our result confirms the only previous measurement \cite{orts_exploring_2006}, but has an uncertainty that is almost nine orders of magnitude lower. This newly achieved level of uncertainty allows benchmarking of predictions from atomic structure calculations. In particular we are now able to resolve contributions from the QED nuclear recoil.\par

Our relativistic CI calculation is supplemented by the separate treatment of the QED effects to the mass shift and to the energy levels. The QED calculation of the mass shift is carried out within the rigorous approach that goes beyond the external field approach of QED and takes into account all orders in the electron-nucleus Coulomb interaction \cite{shabaev_qed_1998}. Details about the theoretical calculations can be found in the Methods section and the Supplementary Material.

Theoretical results for the individual effects contributing to the isotope shift are presented in Table~\ref{tab:tot}. It is seen that our calculations are in good agreement with but more accurate than the previous computations reported in the literature. The accuracy of the total theoretical value for the isotope shift was improved by a factor of three as compared to the previous best calculation \cite{zubova_isotope_2016}. We observe that the QED part of the mass shift is comparable to the field shift and thus cannot be neglected in the theoretical description of the isotope shift as has been customary in the literature previously. The experimental result is in excellent agreement with the theoretical prediction. It should be mentioned that the calculations and experiments were carried out blind to one another, preventing any bias. The comparison of the experimental and theoretical results tests the QED mass shift on the level of 31\%. To the best of our knowledge, this is the first experimental verification of the QED nuclear recoil effect in a many-electron system, since all previous tests of QED were performed within the static external-field regime.

Data taken during the clock comparisons allows the determination of atomic properties as described in the Methods section. We obtain a quadrupole moment of the \pthreehalves{} state of -0.02(9)~$ea_{0}^{2}$ for \Arforty{} and -0.03(14)~$ea_{0}^{2}$ for \Arthirtysix{}, where $e$ is the elementary charge and $a_{0}$ the Bohr radius. Both results confirm the predicted small value of 0.0235\,$ea_{0}^{2}$ \cite{yu_investigating_2019, robert_priv_2021}. The ratio between the $g$-factors of the excited and ground states $r = g_{3/2}/g_{1/2}$ was determined to be $r($\Arforty{}$)=2.007\,514\,2(25)$ for \Arforty{}. With the previously measured $g_{1/2}$ from Ref.~\cite{arapoglou_g-factor_2019}, which has parts-per-billion uncertainty, we obtain $g_{3/2} = 1.332\,283\,7(17)$, being in good agreement with our previous measurement \cite{micke_coherent_2020}. The here-achieved fourfold reduced uncertainty brings it to the level of a theoretical prediction of $g_{3/2}^\textrm{th}=1.332\,282\,5(14)$, and confirms it \cite{agababaev_g_2019}.\par

For \Arthirtysix{}, we determine a $g$-factor ratio of $r($\Arthirtysix{}$)=2.007\,515\,4(32)$, which is not significantly different from the $r($\Arforty{}$)$ result. No precise experimental value is available for $g_{1/2}$ for this isotope. The difference $r($\Arthirtysix{}$) - r($\Arforty{}$) = 1.2(36) \times 10^{-6}$ is consistent with the theoretical prediction of $2.2 \times 10^{-6}$ \cite{agababaev_ground-state_2018,agababaev_g_2019}.\par

\begin{table}[t]
    \caption{\textbf{Theoretical and experimental values for the isotope shift} }
    \begin{tabular*}{\textwidth}{l @{\extracolsep{\fill}} w{2.7}rr}
        \hline
        \multicolumn{1}{l}{}
            & \multicolumn{1}{c}{This work}
                & \multicolumn{1}{c}{CI-DFS \cite{zubova_isotope_2016}}
                    & \multicolumn{1}{c}{MCDF \cite{naze_isotope_2014}}\\
        \hline
        Mass shift, relativistic  & 1906.45\,(80)  &   $1920\; \ \ \ \ \ $   &  $1915$  \\
        Mass shift, QED           &  -13.0\,(36)  &    $-17\,(11)$              \\
        Field shift, relativistic &  -15.9\,(15)   &    $-16\,(1)\, \ $  &  $-16$   \\
        Field shift, QED          &   -0.08        &                             \\
        Total theory              & 1877.5\,(40)  &   $1886\,(11)$   &  $1899$ \\
        \hline
        \hline
        Experiment, this work      &  1878.110\,532\,51(11) \\
        Experiment \cite{orts_exploring_2006} &  \multicolumn{1}{c}{$1895(93)\qquad\qquad\!\!$} \\
        \hline
    \end{tabular*}
    \caption*{The various contributions to the isotope shift $\nu$(\Arforty{}) - $\nu$(\Arthirtysix{}) are discussed in the main text, with further details in the Methods section and Supplementary Material. The total calculated shift is in excellent agreement with the experimental value. All shifts are given in MHz. 
    \label{tab:tot}
    }
\end{table}

\section{Conclusions}
The first operation and evaluation of an optical atomic clock based on a highly charged ion was presented. The achieved performance confirms the applicability of HCI in optical clocks. The frequency instability of the HCI clock can be greatly improved by choosing an HCI species with a longer excited state lifetime, such as Ni$^{12+}$, Pd$^{12+}$ or Pr$^{9+}$ \cite{yu_selected_2018,bekker_detection_2019}. This will allow HCI-based clocks to compete with the best available optical atomic clocks. \par
The potential of HCI to contribute to fundamental physics was demonstrated by experimentally confirming QED nuclear recoil contributions to the isotope shift between \Arforty{} and \Arthirtysix{}. This was possible since the properties of few-electron HCI are well calculable using \textit{ab initio} atomic structure methods. Following this work, isotope shift measurements of optical clock transitions can now be extended to very different charge states of the same element and thus contribute to the search for hypothetical fifth forces \cite{berengut_generalized_2020, rehbehn_sensitivity_2021, kozlov_hci}.
Frequency comparisons between HCI clocks and other optical clocks as demonstrated here can also be employed to probe for beyond-the-standard-model physics. In particular, HCI optical clock transitions have the largest known sensitivity to both a change in the fine-structure constant and to ultralight scalar dark matter of all known atomic systems \cite{schiller_hydrogenlike_2007,berengut_enhanced_2010, berengut_optical_2012, dzuba_highly_2015, porsev_optical_2020, safronova_search_2018}.

\section{Methods}
\subsection{Loading of the different isotopes}

Highly charged argon ions are produced in a miniature EBIT \cite{micke_heidelberg_2018} by injecting gas (here $^{36}$Ar or $^{40}$Ar) through a differentially-pumped injection system into the trap region, where the atoms are ionised by the electron beam. $^{40}$Ar has a natural abundance of $99.6\,\%$, and hence no additional purification stages are required within the EBIT and beamline when working with this isotope. The second-most abundant isotope is $^{36}$Ar, with a natural abundance of only $0.33\,\%$. We use an enriched sample of this isotope with close to $100\,\%$ purity. In order to minimise the necessary quantity of the enriched gas, a dedicated injection system with small dead volume was built.\par
The EBIT operating parameters and breeding time are optimised for maximum yield of \Ar{}. After a short breeding time of a few hundred milliseconds, the HCIs are ejected from the EBIT and launched into an ion-optical beamline connecting it to the Paul trap \cite{schmoger_coulomb_2015, micke_coherent_2020}. The velocities of the different charge states and hence their times-of-flight through the beamline depend on their charge-to-mass ratios. We filter out unwanted charge states in the beamline with an ion-optical element referred to as a `gate' electrode normally held at a high potential to reject the HCIs. It is pulsed to a lower value at the arrival time of the desired charge state, allowing it to pass. The lower mass of $^{36}$Ar necessitates minor changes to the switching times of the beamline electrodes compared to \Arforty{}.

\subsection{Clock and logic lasers}

A small fraction of the 441~nm light generated by frequency doubling of the 882~nm laser light is used to injection-lock a Fabry-P\'erot laser diode with a free-running wavelength of 440~nm at $30^{\circ}$C, which provides the higher-power beams required for fast quantum logic and algorithmic cooling \cite{king_algorithmic_2021} operations. The entire clock and logic laser setup is enclosed in a housing to avoid air flow and temperature variations, allowing operation over several days without adjustments.\par

The clock laser beams are delivered in the radial direction of the trap, perpendicular to the symmetry axis. This has two advantages over our original configuration \cite{micke_coherent_2020}, where the beam entered along the symmetry axis: firstly, the motional modes that the laser can be coupled to have a higher frequency and thus the motional sidebands are better separated from the carrier transition, and yield a smaller Lamb-Dicke parameter. This suppresses line-pulling effects on the carrier from the sideband transitions. Secondly, it is desirable to alternate the direction of interrogation between two counter-propagating directions to eliminate any potential first-order Doppler shift. This is more difficult along the trap axis owing to the presence of the 1.2~m-long ion-optical beamline. The QLS beam still enters along the trap axis to ensure strong coupling to the axial sidebands used in this method.\par

The average of the frequencies of the eight servos steer the clock output by adjusting the offset of the clock laser to the optical frequency comb mode. This decouples the clock laser from the ultrastable \SI{1.5}{\micro\meter} laser on timescales beyond the servo time constant and ensures that the laser frequency measured using the comb is a real-time reflection of the transition frequency with the linear Zeeman, linear Doppler, and quadrupole shifts removed. Only a constant frequency offset remains, which depends on the drive frequencies of the various AOMs. This procedure makes it unnecessary to carry out synchronised measurements of several radiofrequency signals such as AOM drive frequencies in the various laboratories.

\subsection{Quadrupole moment of \pthreehalves{} state}
States with $J\ge1$ such as the excited \pthreehalves{} state in \Ar{} possess electric quadrupole moments $\Theta$. Electric field gradients experienced by the HCI will therefore lead to quadrupole shifts of its sublevels that depend on $\left\lvert m_J\right\rvert ^2$ \cite{itano_external-field_2000}. For a two-ion crystal aligned along the symmetry axis of a linear Paul trap, the d.c. potential $U$ used for axial confinement has a gradient at the position of the HCI. For our trap geometry, the axial potential is described by:
\begin{equation}
    U = \frac{1}{2} \frac{d E_z}{d z} (x^2 - z^2)
\end{equation}
where $d E_z / d z$ is the electric field gradient along the axial symmetry axis $z$ of the trap. The $x$ axis is defined along the two rf blade electrodes \cite{leopold_cryogenic_2019}. The lack of $y$-dependence is specific to the geometry of our trap and is confirmed by simulations and experiments. The gradient $d E_z / d z$ can be determined by measuring the single- or two-ion motional frequencies in the $z$-direction. A second contribution arising from the charge of the \Be{} ion \cite{akerman_atomic_2018} is smaller by a factor of seven compared to that of the trapping field, due to its large separation of \SI{21}{\micro\meter} from the \Ar{} ion.

This electric field gradient of the trap shifts the Zeeman sublevels of the excited state according to the relation
\begin{equation}
	\Delta \nu_\text{qs} = \frac{\Theta}{4h}  \frac{d E_z}{d z} \frac{J\left(J+1\right) - 3m_J^2 }{J \left( 2J-1 \right)} \left(3 \cos^2 \beta -1 - \epsilon \sin^2 \beta \cos (2\alpha) \right),
	\label{eq:QS}
\end{equation}
where $\beta = 30(5) ^\circ$ is the fixed angle between the electric field gradient and the quantisation axis in our setup, $\alpha = 45(5)^\circ$ the angle between the $x$-axis and the plane defined by the $z$-axis and our quantisation axis, and $\epsilon = 1$ accounts for the breaking of cylindrical symmetry. The shift is averaged in real time by choosing to operate with servos that sample all of the Zeeman substates of the excited state. Nevertheless, its value can be determined by comparing the mean frequency of pairs of Zeeman components that share the same value for $\left\lvert m_J \right\rvert ^2$.

\subsection{$g$-factor of \pthreehalves{} state}
The ratio $r$ between the Land\'e $g$-factors of the excited $g_{3/2}$ and the $g_{1/2}$ ground state can be determined from the ratio $\rho$ of the splitting between the outer and inner Zeeman component pairs:
\begin{equation}
	r=\frac{g_{3/2}}{g_{1/2}} = \frac{1 - \rho}{3 - \rho},
\end{equation}
with $\rho = (\nu_{4} - \nu_{1})/(\nu_{3} - \nu_{2})$, where the labelling of the four transitions follows that in Fig.~\ref{fig:transitions} in the main section. We calculate $r$ from servo data on a cycle-by-cycle basis in order to avoid degradation of the measurement due to drifts in the magnetic field during the measurement period.\par 
The measured ratio $r$ is also affected by the oscillating magnetic field from the rf drive of the trap, which couples the Zeeman substates of each manifold \cite{Gan2018,arnold_precision_2020}. Previously, its mean-square amplitude along the quantisation axis of our trap has been bounded to be smaller than \SI{3}{\micro\tesla\squared} at drive amplitudes twice as high as in the present work \cite{leopold_cryogenic_2019}. For a conservative estimate of this contribution, we use this value and assume based on the geometry of our trap that its transverse components have the same magnitude. This leads to a fractional shift bounded to $\vert\Delta r/r\vert < 7\times10^{-7}$, with this value taken fully as an additional uncertainty.

\subsection{Data processing}
Spurious data was identified and removed in various ways. Firstly, cycle slips on the tracking oscillators used by the \Ar{} experiment were checked by having two oscillators with asymmetrically detuned free-running frequencies and with slightly different loop gains tracking each beat signal. Both are counted and typically agree at the mHz-level for sampling times of 1~s used for all frequency counting processes. To detect the cycle slips, we apply a threshold value of 0.1~Hz difference between the two counters, as a cycle slip of $2\pi$ during the sampling time of 1\,s would result in a 1\,Hz jump. For the other beats involved in the frequency comparison, only a single counter channel was used. In these cases, deviations were found by applying a threshold of 0.1~Hz from the nominal, fixed beat frequency, which is between 5 and 100 times larger than the typical maximum fluctuations of the various beat frequencies. Secondly, any sudden glitches in the \Ar{} clock laser frequency caused by a momentary loss of the phase lock to the laser stabilised to the Si2 cavity \cite{matei_1.5_2017} were identified.\par
In all cases described above, the faulty data was removed, along with 30~s of data afterwards. In addition, 5~s of data was removed before the faulty period to allow for any offset in the synchronisation between the frequency counters in the different laboratories.\par
Further data was excluded when one or more of the eight clock servos temporarily lost lock, for example after an unusually rapid change in the local magnetic field. Data in the vicinity of unusual changes in the magnetic field was removed, even if the servos remained in lock throughout.\par
To exclude possible errors in the data analysis, at least two independent blind evaluations of each dataset were carried out. In all cases, they agreed well within the measurement uncertainties.

\subsection{Theory of the isotope shift}

The isotope-dependent part of an atomic transition energy can be represented as a sum of the so-called mass-shift ($E_{\rm ms}$) and field-shift ($E_{\rm fs}$) terms,
\begin{eqnarray}\label{eq:iso}
E_{\rm iso} = E_{\rm ms} + E_{\rm fs} = \frac{m_e}{M}\, K +  \Big(\frac{R}{\lambdabar_C}\Big)^2\,F\,,
\end{eqnarray}
where $m_e$ is the electron mass, $M$ is the nuclear mass, $R = \big< r^2 \big>^{1/2}$ is the
root-mean-square (rms) radius of the nuclear charge distribution, and $\lambdabar_C = \lambda_C/(2\pi)$ is the reduced Compton wavelength. $K$ and $F$ are the mass- and the field-shift constants, respectively; they depend on the electronic structure of the atom but not on the nuclear properties of the isotope.

The relativistic mass-shift constant is induced \cite{shabaev_qed_1998} by the expectation value of the relativistic recoil operator, $K = \big< \widetilde{H}_{\rm rec} \big>$, with
\begin{eqnarray}
\widetilde{H}_{\rm rec} = \frac12\,\sum_{ik} \bigg[ \bfp_i\cdot\bfp_k
-\frac{Z\alpha}{r_i} \Big( \balpha_i + \frac{(\balpha_i\cdot\bfr_i)\,\bfr_i}{r_i^2} \Big)
\cdot\bfp_k \bigg]\,.
\end{eqnarray}
Here, the indices $i$ and $k$ numerate the electrons, $\bfp$ and $\bfr$ are the momentum and coordinate operators, and $\balpha$ is the vector of Dirac matrices. The relativistic mass shift of the \phalf{} $\rightarrow$ \pthreehalves{} transition in B-like argon was calculated by the configuration-interaction (CI) method, implemented in our previous works \cite{yerokhin_nonlinear_2020}. In our calculation, we account for a strong mixing of the $1s^22s^22p_j$ states with the closely lying excited states by treating all possible $1s^22l2l'2l''$ configurations as multiple reference states, where $l$, $l'$, and $l''$ denote angular momenta of electron orbitals. The CI expansion of our large-scale computation included single, double, triple, and quadruple
excitations from the multiple reference states specified above. \par    

In order to obtain accurate results for the mass-shift constant $K$, the relativistic CI treatment needs to be supplemented by a separate treatment of QED nuclear recoil effects. The underlying theory was developed in Ref.~\cite{shabaev_qed_1998}. In the present work, we calculate the QED recoil effect within the independent-electron approximation, for which the one-electron wave functions are obtained by solving the Dirac equation with a localised Dirac-Fock potential. Furthermore, we account for the correction arising from the QED shifts of energy levels. Because of a large mixing between closely lying states, the QED energy shifts alter the mixing coefficients, and thus induce a significant contribution to the mass-shift constant. This correction was calculated by including the effective QED operator \cite{shabaev_model_2013,shabaev_qedmod_2015} into the Hamiltonian matrix of the CI procedure.\par%

The relativistic value for the field-shift constant $F$ is obtained as an expectation value of the derivative of the nuclear binding potential $V_{\rm nuc}$ over the square of the nuclear charge radius $R^2$,
\begin{eqnarray}
F =  \Big< V_{\rm FS} \Big>  = \Big< \sum_i \frac{d V_{\rm nuc}(r_i)}{d (R/\lambdabar_C)^2} \Big>\,. 
\end{eqnarray}
We calculate the constant $F$ by averaging the operator $V_{\rm FS}$ with the CI wave functions obtained in the same way as for the mass shift. The field shift of the \phalf{} $\rightarrow$ \pthreehalves{} transition energy in \Ar{} is much smaller than the mass shift. The reason is that the overlap of the $2p$ one-electron orbitals with the nucleus is very small. Thus, the dominant effect comes from mixing with the $1s$ core orbital, which is suppressed by the parameter $1/Z$. Because of its smallness, the field shift does not introduce any contribution to the uncertainty budget of the total isotope shift in the case under consideration.\par

The relativistic value of the field-shift constant needs to be supplemented by an estimate of the leading QED effects. We find that the leading contribution comes from the QED shifts of energy levels. This effect is enhanced due to strong mixing of the reference states with the closely-lying virtual excited states. We calculate the QED correction due to energy shifts by including the effective QED operator \cite{shabaev_model_2013,shabaev_qedmod_2015} into the Hamiltonian matrix of the CI method, and repeating our calculations with and without the QED addition. The second largest QED effect is the pure radiative finite-nuclear-size correction. Following Ref.~\cite{zubova_isotope_2016}, we estimate it by rescaling the known hydrogenic result for the $1s$ state from Ref.~\cite{yerokhin_nuclear-size_2011}.\par %

We summarise our calculations in Table~\ref{tab:tot} in the main section. The nuclear parameters used in the calculation are: $M_{40}/m_e = 72828.99631$, $M_{36}/m_e = 65546.85208$, $R_{40} = 3.4274(26)$~fm, $R_{36} = 3.3905(23)$~fm. The charge radii are taken from Ref.~\cite{angeli_table_2013}. The quoted error of the theoretical prediction includes 1.5~MHz uncertainty due to the experimental values of the nuclear radii. The nuclear masses are calculated by subtracting the mass of the electrons, and binding energies from the atomic masses listed in Ref.~\cite{wang_ame2012_2012}. Further details about these calculations are given in the Supplementary Material.

\vspace{2 mm}

\textbf{Data availability} The traces of the frequency ratios are available at \url{https://doi.org/10.5281/zenodo.6901524}. Additional datasets generated and analysed during this study are available from the corresponding author upon request.

\textbf{Code availability} All code that has been used to generate or analyse data during this study are available from the corresponding author upon request.

\textbf{Acknowledgments}

The authors would like to thank Lisa Schm\"oger, Maria Schwarz, and Julian Stark for early contributions to the experimental apparatus, Thomas Legero for his contributions to the frequency stabilisation of the HCI spectroscopy laser, Helen Margolis for discussions about the analysis of the frequency data, and Fabian Wolf for comments on the manuscript. A.S. and V.A.Y. thank I. I. Tupitsyn for discussions. The project was supported by the Physikalisch-Technische Bundesanstalt, the Max-Planck Society, the Max-Planck–Riken–PTB–Center for Time, Constants and Fundamental Symmetries, and the Deutsche Forschungsgemeinschaft (DFG, German Research Foundation) through SCHM2678/5-1, SU 658/4-2, the collaborative research centres SFB 1225 ISOQUANT and SFB 1227 DQ-mat, and under Germany’s Excellence Strategy – EXC-2123 QuantumFrontiers – 390837967. These projects 17FUN07 CC4C and 20FUN01 TSCAC have received funding from the EMPIR programme co-financed by the Participating States and from the European Union’s Horizon 2020 research and innovation programme. This project has received funding from the European Research Council (ERC) under the European Union’s Horizon 2020 research and innovation programme (grant agreement No 101019987). S.A.K. acknowledges financial support from the Alexander von Humboldt Foundation.

\textbf{Author Contributions} S.A.K., L.J.S., P.M., T.L., E.B., J.R.C.L.-U., and P.O.S. developed the experimental setup. S.A.K., L.J.S., P.M., A.W., R.L., and N.H. carried out the experiments. S.A.K., L.J.S., A.W., and E.B. analysed the data. J.R.C.L.-U. and P.O.S. conceived and supervised the study. A.S. and V.A.Y. performed the theoretical calculations. S.A.K., L.J.S., A.S., and P.O.S. wrote the initial manuscript with contributions from P.M. and J.R.C.L.-U.. All authors discussed the results and reviewed the manuscript.

\textbf{Author Information} The authors declare no competing interests. Reprints and permissions information is available at www.nature.com/reprints. Supplementary information is available for this paper. Correspondence should be addressed to L.J.S. (lukas.spiess@quantummetrology.de) or P.O.S. (piet.schmidt@quantummetrology.de).

\clearpage

\textbf{Extended Data Table 1 | Measured frequency ratios and absolute frequencies}

\begin{center}
\includegraphics[scale=1.5, trim=6cm 14.3cm 6cm 13cm,clip]{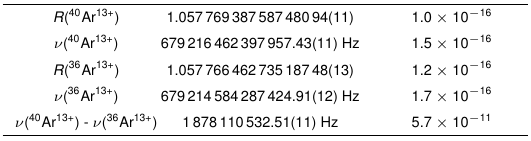}
\end{center}

Optical frequency ratios $R$($^{X}$Ar$^{13+}$) $=\nu$($^{X}$Ar$^{13+}$) / $\nu$(\Yb{} $E3$), derived transition frequencies $\nu$($^{X}$Ar$^{13+}$), resulting isotope shift $\nu$(\Arforty{}) - $\nu$(\Arthirtysix{}), and total relative uncertainties of each of the measurements are given.

\clearpage

\textbf{Extended Data Table 2 | Investigated systematic shifts ($\Delta\nu$) and corresponding 1-$\sigma$ uncertainties ($\sigma$) for the \Ar{} clock}

\begin{center}
\includegraphics[scale=1.5, trim=5.2cm 12.7cm 5.2cm 10.9cm,clip]{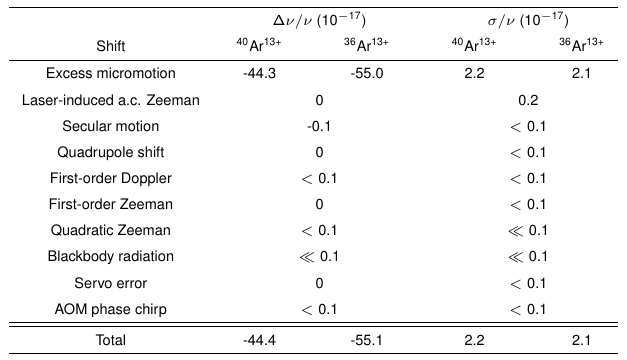}
\end{center}

Values are given relative to the transition frequency $\nu$. A detailed analysis is given in the Supplementary Material.


\begin{thebibliography}{62}
\ifx \bisbn   \undefined \def \bisbn  #1{ISBN #1}\fi
\ifx \binits  \undefined \def \binits#1{#1}\fi
\ifx \bauthor  \undefined \def \bauthor#1{#1}\fi
\ifx \batitle  \undefined \def \batitle#1{#1}\fi
\ifx \bjtitle  \undefined \def \bjtitle#1{#1}\fi
\ifx \bvolume  \undefined \def \bvolume#1{\textbf{#1}}\fi
\ifx \byear  \undefined \def \byear#1{#1}\fi
\ifx \bissue  \undefined \def \bissue#1{#1}\fi
\ifx \bfpage  \undefined \def \bfpage#1{#1}\fi
\ifx \blpage  \undefined \def \blpage #1{#1}\fi
\ifx \burl  \undefined \def \burl#1{\textsf{#1}}\fi
\ifx \doiurl  \undefined \def \doiurl#1{\url{https://doi.org/#1}}\fi
\ifx \betal  \undefined \def \betal{\textit{et al.}}\fi
\ifx \binstitute  \undefined \def \binstitute#1{#1}\fi
\ifx \binstitutionaled  \undefined \def \binstitutionaled#1{#1}\fi
\ifx \bctitle  \undefined \def \bctitle#1{#1}\fi
\ifx \beditor  \undefined \def \beditor#1{#1}\fi
\ifx \bpublisher  \undefined \def \bpublisher#1{#1}\fi
\ifx \bbtitle  \undefined \def \bbtitle#1{#1}\fi
\ifx \bedition  \undefined \def \bedition#1{#1}\fi
\ifx \bseriesno  \undefined \def \bseriesno#1{#1}\fi
\ifx \blocation  \undefined \def \blocation#1{#1}\fi
\ifx \bsertitle  \undefined \def \bsertitle#1{#1}\fi
\ifx \bsnm \undefined \def \bsnm#1{#1}\fi
\ifx \bsuffix \undefined \def \bsuffix#1{#1}\fi
\ifx \bparticle \undefined \def \bparticle#1{#1}\fi
\ifx \barticle \undefined \def \barticle#1{#1}\fi
\ifx \bconfdate \undefined \def \bconfdate #1{#1}\fi
\ifx \botherref \undefined \def \botherref #1{#1}\fi
\ifx \url \undefined \def \url#1{\textsf{#1}}\fi
\ifx \bchapter \undefined \def \bchapter#1{#1}\fi
\ifx \bbook \undefined \def \bbook#1{#1}\fi
\ifx \bcomment \undefined \def \bcomment#1{#1}\fi
\ifx \oauthor \undefined \def \oauthor#1{#1}\fi
\ifx \citeauthoryear \undefined \def \citeauthoryear#1{#1}\fi
\ifx \endbibitem  \undefined \def \endbibitem {}\fi
\ifx \bconflocation  \undefined \def \bconflocation#1{#1}\fi
\ifx \arxivurl  \undefined \def \arxivurl#1{\textsf{#1}}\fi
\csname PreBibitemsHook\endcsname

\bibitem{ludlow_optical_2015}
\begin{barticle}
\bauthor{\bsnm{Ludlow}, \binits{A.D.}},
\bauthor{\bsnm{Boyd}, \binits{M.M.}},
\bauthor{\bsnm{Ye}, \binits{J.}},
\bauthor{\bsnm{Peik}, \binits{E.}},
\bauthor{\bsnm{Schmidt}, \binits{P.O.}}:
\batitle{Optical atomic clocks}.
\bjtitle{Rev. Mod. Phys.}
\bvolume{87}(\bissue{2}),
\bfpage{637}--\blpage{701}
(\byear{2015}).
\doiurl{10.1103/RevModPhys.87.637}.
\end{barticle}
\endbibitem

\bibitem{safronova_search_2018}
\begin{barticle}
\bauthor{\bsnm{Safronova}, \binits{M.S.}},
\bauthor{\bsnm{Budker}, \binits{D.}},
\bauthor{\bsnm{DeMille}, \binits{D.}},
\bauthor{\bsnm{Kimball}, \binits{D.F.J.}},
\bauthor{\bsnm{Derevianko}, \binits{A.}},
\bauthor{\bsnm{Clark}, \binits{C.W.}}:
\batitle{Search for new physics with atoms and molecules}.
\bjtitle{Rev. Mod. Phys.}
\bvolume{90}(\bissue{2}),
\bfpage{025008}
(\byear{2018}).
\doiurl{10.1103/RevModPhys.90.025008}.
\end{barticle}
\endbibitem

\bibitem{mehlstaubler_atomic_2018}
\begin{barticle}
\bauthor{\bsnm{Mehlstäubler}, \binits{T.E.}},
\bauthor{\bsnm{Grosche}, \binits{G.}},
\bauthor{\bsnm{Lisdat}, \binits{C.}},
\bauthor{\bsnm{Schmidt}, \binits{P.O.}},
\bauthor{\bsnm{Denker}, \binits{H.}}:
\batitle{Atomic clocks for geodesy}.
\bjtitle{Rep. Prog. Phys.}
\bvolume{81}(\bissue{6}),
\bfpage{064401}
(\byear{2018}).
\doiurl{10.1088/1361-6633/aab409}.
\end{barticle}
\endbibitem

\bibitem{kozlov_hci}
\begin{barticle}
\bauthor{\bsnm{Kozlov}, \binits{M.G.}},
\bauthor{\bsnm{Safronova}, \binits{M.S.}},
\bauthor{\bsnm{Crespo~L\'opez-Urrutia}, \binits{J.R.}},
\bauthor{\bsnm{Schmidt}, \binits{P.O.}}:
\batitle{Highly charged ions: Optical clocks and applications in fundamental
  physics}.
\bjtitle{Rev. Mod. Phys.}
\bvolume{90},
\bfpage{045005}
(\byear{2018}).
\doiurl{10.1103/RevModPhys.90.045005}
\end{barticle}
\endbibitem

\bibitem{schiller_hydrogenlike_2007}
\begin{barticle}
\bauthor{\bsnm{Schiller}, \binits{S.}}:
\batitle{Hydrogenlike {Highly} {Charged} {Ions} for {Tests} of the {Time}
  {Independence} of {Fundamental} {Constants}}.
\bjtitle{Phys. Rev. Lett.}
\bvolume{98}(\bissue{18}),
\bfpage{180801}
(\byear{2007}).
\doiurl{10.1103/PhysRevLett.98.180801}
\end{barticle}
\endbibitem

\bibitem{berengut_enhanced_2010}
\begin{barticle}
\bauthor{\bsnm{Berengut}, \binits{J.}},
\bauthor{\bsnm{Dzuba}, \binits{V.}},
\bauthor{\bsnm{Flambaum}, \binits{V.}}:
\batitle{Enhanced {Laboratory} {Sensitivity} to {Variation} of the
  {Fine}-{Structure} {Constant} using {Highly} {Charged} {Ions}}.
\bjtitle{Phys. Rev. Lett.}
\bvolume{105}(\bissue{12}),
\bfpage{120801}
(\byear{2010}).
\doiurl{10.1103/PhysRevLett.105.120801}
\end{barticle}
\endbibitem

\bibitem{berengut_optical_2012}
\begin{barticle}
\bauthor{\bsnm{Berengut}, \binits{J.C.}},
\bauthor{\bsnm{Dzuba}, \binits{V.A.}},
\bauthor{\bsnm{Flambaum}, \binits{V.V.}},
\bauthor{\bsnm{Ong}, \binits{A.}}:
\batitle{Optical {Transitions} in {Highly} {Charged} {Californium} {Ions} with
  {High} {Sensitivity} to {Variation} of the {Fine}-{Structure} {Constant}}.
\bjtitle{Phys. Rev. Lett.}
\bvolume{109}(\bissue{7}),
\bfpage{070802}
(\byear{2012}).
\doiurl{10.1103/PhysRevLett.109.070802}.
\end{barticle}
\endbibitem

\bibitem{derevianko_highly_2012}
\begin{barticle}
\bauthor{\bsnm{Derevianko}, \binits{A.}},
\bauthor{\bsnm{Dzuba}, \binits{V.A.}},
\bauthor{\bsnm{Flambaum}, \binits{V.V.}}:
\batitle{Highly {Charged} {Ions} as a {Basis} of {Optical} {Atomic} {Clockwork}
  of {Exceptional} {Accuracy}}.
\bjtitle{Phys. Rev. Lett.}
\bvolume{109}(\bissue{18}),
\bfpage{180801}
(\byear{2012}).
\doiurl{10.1103/PhysRevLett.109.180801}.
\end{barticle}
\endbibitem

\bibitem{safronova_highly_2014-1}
\begin{barticle}
\bauthor{\bsnm{Safronova}, \binits{M.S.}},
\bauthor{\bsnm{Dzuba}, \binits{V.A.}},
\bauthor{\bsnm{Flambaum}, \binits{V.V.}},
\bauthor{\bsnm{Safronova}, \binits{U.I.}},
\bauthor{\bsnm{Porsev}, \binits{S.G.}},
\bauthor{\bsnm{Kozlov}, \binits{M.G.}}:
\batitle{Highly charged {Ag}-like and {In}-like ions for the development of
  atomic clocks and the search for α variation}.
\bjtitle{Phys. Rev. A}
\bvolume{90}(\bissue{4}),
\bfpage{042513}
(\byear{2014}).
\doiurl{10.1103/PhysRevA.90.042513}.
\end{barticle}
\endbibitem

\bibitem{yudin_magnetic-dipole_2014}
\begin{barticle}
\bauthor{\bsnm{Yudin}, \binits{V.I.}},
\bauthor{\bsnm{Taichenachev}, \binits{A.V.}},
\bauthor{\bsnm{Derevianko}, \binits{A.}}:
\batitle{Magnetic-{Dipole} {Transitions} in {Highly} {Charged} {Ions} as a
  {Basis} of {Ultraprecise} {Optical} {Clocks}}.
\bjtitle{Phys. Rev. Lett.}
\bvolume{113}(\bissue{23}),
\bfpage{233003}
(\byear{2014}).
\doiurl{10.1103/PhysRevLett.113.233003}.
\end{barticle}
\endbibitem

\bibitem{beloy_quadruply_2020}
\begin{barticle}
\bauthor{\bsnm{Beloy}, \binits{K.}},
\bauthor{\bsnm{Dzuba}, \binits{V.A.}},
\bauthor{\bsnm{Brewer}, \binits{S.M.}}:
\batitle{Quadruply {Ionized} {Barium} as a {Candidate} for a {High}-{Accuracy}
  {Optical} {Clock}}.
\bjtitle{Phys. Rev. Lett.}
\bvolume{125}(\bissue{17}),
\bfpage{173002}
(\byear{2020}).
\doiurl{10.1103/PhysRevLett.125.173002}.
\end{barticle}
\endbibitem

\bibitem{egl_application_2019}
\begin{barticle}
\bauthor{\bsnm{Egl}, \binits{A.}},
\bauthor{\bsnm{Arapoglou}, \binits{I.}},
\bauthor{\bsnm{H\"ocker}, \binits{M.}},
\bauthor{\bsnm{K\"onig}, \binits{K.}},
\bauthor{\bsnm{Ratajczyk}, \binits{T.}},
\bauthor{\bsnm{Sailer}, \binits{T.}},
\bauthor{\bsnm{Tu}, \binits{B.}},
\bauthor{\bsnm{Weigel}, \binits{A.}},
\bauthor{\bsnm{Blaum}, \binits{K.}},
\bauthor{\bsnm{N\"ortersh\"auser}, \binits{W.}},
\bauthor{\bsnm{Sturm}, \binits{S.}}:
\batitle{{Application} of the {Continuous} {Stern}-{Gerlach} {Effect} for
  {Laser} {Spectroscopy} of the {$^{40}{\mathrm{Ar}}^{13+}$} {Fine} {Structure}
  in a {Penning} {Trap}}.
\bjtitle{Phys. Rev. Lett.}
\bvolume{123},
\bfpage{123001}
(\byear{2019}).
\doiurl{10.1103/PhysRevLett.123.123001}
\end{barticle}
\endbibitem

\bibitem{orts_exploring_2006}
\begin{barticle}
\bauthor{\bsnm{Orts}, \binits{R.S.}},
\bauthor{\bsnm{Harman}, \binits{Z.}},
\bauthor{\bsnm{López-Urrutia}, \binits{J.R.C.}},
\bauthor{\bsnm{Artemyev}, \binits{A.N.}},
\bauthor{\bsnm{Bruhns}, \binits{H.}},
\bauthor{\bsnm{Martínez}, \binits{A.J.G.}},
\bauthor{\bsnm{Jentschura}, \binits{U.D.}},
\bauthor{\bsnm{Keitel}, \binits{C.H.}},
\bauthor{\bsnm{Lapierre}, \binits{A.}},
\bauthor{\bsnm{Mironov}, \binits{V.}},
\bauthor{\bsnm{Shabaev}, \binits{V.M.}},
\bauthor{\bsnm{Tawara}, \binits{H.}},
\bauthor{\bsnm{Tupitsyn}, \binits{I.I.}},
\bauthor{\bsnm{Ullrich}, \binits{J.}},
\bauthor{\bsnm{Volotka}, \binits{A.V.}}:
\batitle{Exploring {Relativistic} {Many}-{Body} {Recoil} {Effects} in {Highly}
  {Charged} {Ions}}.
\bjtitle{Phys. Rev. Lett.}
\bvolume{97}(\bissue{10}),
\bfpage{103002}
(\byear{2006}).
\doiurl{10.1103/PhysRevLett.97.103002}.
\end{barticle}
\endbibitem

\bibitem{zubova_isotope_2016}
\begin{barticle}
\bauthor{\bsnm{Zubova}, \binits{N.A.}},
\bauthor{\bsnm{Malyshev}, \binits{A.V.}},
\bauthor{\bsnm{Tupitsyn}, \binits{I.I.}},
\bauthor{\bsnm{Shabaev}, \binits{V.M.}},
\bauthor{\bsnm{Kozhedub}, \binits{Y.S.}},
\bauthor{\bsnm{Plunien}, \binits{G.}},
\bauthor{\bsnm{Brandau}, \binits{C.}},
\bauthor{\bsnm{Stöhlker}, \binits{T.}}:
\batitle{Isotope shifts of the 2p$_{\textrm{3/2}}$-2p$_{\textrm{1/2}}$
  transition in {B}-like ions}.
\bjtitle{Phys. Rev. A}
\bvolume{93}(\bissue{5}),
\bfpage{052502}
(\byear{2016}).
\doiurl{10.1103/PhysRevA.93.052502}.
\end{barticle}
\endbibitem

\bibitem{gillaspy_highly_2001}
\begin{barticle}
\bauthor{\bsnm{Gillaspy}, \binits{J.D.}}:
\batitle{Highly charged ions}.
\bjtitle{J. Phys. B}
\bvolume{34},
\bfpage{93}--\blpage{130}
(\byear{2001}).
\doiurl{10.1088/0953-4075/34/19/201}.
\end{barticle}
\endbibitem

\bibitem{arnold_blackbody_2018}
\begin{barticle}
\bauthor{\bsnm{Arnold}, \binits{K.J.}},
\bauthor{\bsnm{Kaewuam}, \binits{R.}},
\bauthor{\bsnm{Roy}, \binits{A.}},
\bauthor{\bsnm{Tan}, \binits{T.R.}},
\bauthor{\bsnm{Barrett}, \binits{M.D.}}:
\batitle{Blackbody radiation shift assessment for a lutetium ion clock}.
\bjtitle{Nat. Commun}
\bvolume{9}(\bissue{1}),
\bfpage{1}--\blpage{6}
(\byear{2018}).
\doiurl{10.1038/s41467-018-04079-x}.
\end{barticle}
\endbibitem

\bibitem{hachisu_trapping_2008}
\begin{barticle}
\bauthor{\bsnm{Hachisu}, \binits{H.}},
\bauthor{\bsnm{Miyagishi}, \binits{K.}},
\bauthor{\bsnm{Porsev}, \binits{S.G.}},
\bauthor{\bsnm{Derevianko}, \binits{A.}},
\bauthor{\bsnm{Ovsiannikov}, \binits{V.D.}},
\bauthor{\bsnm{Pal’Chikov}, \binits{V.G.}},
\bauthor{\bsnm{Takamoto}, \binits{M.}},
\bauthor{\bsnm{Katori}, \binits{H.}}:
\batitle{Trapping of neutral mercury atoms and prospects for optical lattice
  clocks}.
\bjtitle{Phys. Rev. Lett.}
\bvolume{100}(\bissue{5}),
\bfpage{53001}
(\byear{2008})
\doiurl{10.1103/PhysRevLett.100.053001}
\end{barticle}
\endbibitem

\bibitem{bieber_studies_1997}
\begin{barticle}
\bauthor{\bsnm{Bieber}, \binits{D.J.}},
\bauthor{\bsnm{Margolis}, \binits{H.S.}},
\bauthor{\bsnm{Oxley}, \binits{P.K.}},
\bauthor{\bsnm{Silver}, \binits{J.D.}}:
\batitle{Studies of magnetic dipole transitions in highly charged argon and
  barium using an electron beam ion trap}.
\bjtitle{Phys. Scr.}
\bvolume{T73},
\bfpage{64}--\blpage{66}
(\byear{1997}).
\doiurl{10.1088/0031-8949/1997/T73/015}.
\end{barticle}
\endbibitem

\bibitem{soria_orts_zeeman_2007}
\begin{barticle}
\bauthor{\bsnm{Soria~Orts}, \binits{R.}},
\bauthor{\bsnm{Crespo~L\'opez-Urrutia}, \binits{J.R.}},
\bauthor{\bsnm{Bruhns}, \binits{H.}},
\bauthor{\bsnm{Gonz\'alez~Mart\'{\i}nez}, \binits{A.J.}},
\bauthor{\bsnm{Harman}, \binits{Z.}},
\bauthor{\bsnm{Jentschura}, \binits{U.D.}},
\bauthor{\bsnm{Keitel}, \binits{C.H.}},
\bauthor{\bsnm{Lapierre}, \binits{A.}},
\bauthor{\bsnm{Tawara}, \binits{H.}},
\bauthor{\bsnm{Tupitsyn}, \binits{I.I.}},
\bauthor{\bsnm{Ullrich}, \binits{J.}},
\bauthor{\bsnm{Volotka}, \binits{A.V.}}:
\batitle{Zeeman splitting and $g$ factor of the $1{s}^{2}2{s}^{2}2p$
  $^{2}{P}_{3/2}$ and $^{2}{P}_{1/2}$ levels in {${\mathrm{Ar}}^{13+}$}}.
\bjtitle{Phys. Rev. A}
\bvolume{76},
\bfpage{052501}
(\byear{2007}).
\doiurl{10.1103/PhysRevA.76.052501}
\end{barticle}
\endbibitem

\bibitem{mackel_laser_2011}
\begin{barticle}
\bauthor{\bsnm{Mäckel}, \binits{V.}},
\bauthor{\bsnm{Klawitter}, \binits{R.}},
\bauthor{\bsnm{Brenner}, \binits{G.}},
\bauthor{\bsnm{Crespo~López-Urrutia}, \binits{J.R.}},
\bauthor{\bsnm{Ullrich}, \binits{J.}}:
\batitle{Laser {Spectroscopy} on {Forbidden} {Transitions} in {Trapped}
  {Highly} {Charged} {Ar}$^{\textrm{13+}}$ {Ions}}.
\bjtitle{Phys. Rev. Lett.}
\bvolume{107}(\bissue{14}),
\bfpage{143002}
(\byear{2011}).
\doiurl{10.1103/PhysRevLett.107.143002}.
\end{barticle}
\endbibitem

\bibitem{brewer_27+_2019}
\begin{barticle}
\bauthor{\bsnm{Brewer}, \binits{S.M.}},
\bauthor{\bsnm{Chen}, \binits{J.-S.}},
\bauthor{\bsnm{Hankin}, \binits{A.M.}},
\bauthor{\bsnm{Clements}, \binits{E.R.}},
\bauthor{\bsnm{Chou}, \binits{C.W.}},
\bauthor{\bsnm{Wineland}, \binits{D.J.}},
\bauthor{\bsnm{Hume}, \binits{D.B.}},
\bauthor{\bsnm{Leibrandt}, \binits{D.R.}}:
\batitle{$^{\textrm{27}}${Al}$^{\textrm{+}}$ {Quantum}-{Logic} {Clock} with a
  {Systematic} {Uncertainty} below 10$^{\textrm{-18}}$}.
\bjtitle{Phys. Rev. Lett.}
\bvolume{123}(\bissue{3}),
\bfpage{033201}
(\byear{2019}).
\doiurl{10.1103/PhysRevLett.123.033201}.
\end{barticle}
\endbibitem

\bibitem{schmoger_coulomb_2015}
\begin{barticle}
\bauthor{\bsnm{Schmöger}, \binits{L.}},
\bauthor{\bsnm{Versolato}, \binits{O.O.}},
\bauthor{\bsnm{Schwarz}, \binits{M.}},
\bauthor{\bsnm{Kohnen}, \binits{M.}},
\bauthor{\bsnm{Windberger}, \binits{A.}},
\bauthor{\bsnm{Piest}, \binits{B.}},
\bauthor{\bsnm{Feuchtenbeiner}, \binits{S.}},
\bauthor{\bsnm{Pedregosa-Gutierrez}, \binits{J.}},
\bauthor{\bsnm{Leopold}, \binits{T.}},
\bauthor{\bsnm{Micke}, \binits{P.}},
\bauthor{\bsnm{Hansen}, \binits{A.K.}},
\bauthor{\bsnm{Baumann}, \binits{T.M.}},
\bauthor{\bsnm{Drewsen}, \binits{M.}},
\bauthor{\bsnm{Ullrich}, \binits{J.}},
\bauthor{\bsnm{Schmidt}, \binits{P.O.}},
\bauthor{\bsnm{López-Urrutia}, \binits{J.R.C.}}:
\batitle{Coulomb crystallization of highly charged ions}.
\bjtitle{Science}
\bvolume{347}(\bissue{6227}),
\bfpage{1233}--\blpage{1236}
(\byear{2015}).
\doiurl{10.1126/science.aaa2960}.
\end{barticle}
\endbibitem

\bibitem{micke_coherent_2020}
\begin{barticle}
\bauthor{\bsnm{Micke}, \binits{P.}},
\bauthor{\bsnm{Leopold}, \binits{T.}},
\bauthor{\bsnm{King}, \binits{S.A.}},
\bauthor{\bsnm{Benkler}, \binits{E.}},
\bauthor{\bsnm{Spieß}, \binits{L.J.}},
\bauthor{\bsnm{Schmöger}, \binits{L.}},
\bauthor{\bsnm{Schwarz}, \binits{M.}},
\bauthor{\bsnm{López-Urrutia}, \binits{J.R.C.}},
\bauthor{\bsnm{Schmidt}, \binits{P.O.}}:
\batitle{Coherent laser spectroscopy of highly charged ions using quantum
  logic}.
\bjtitle{Nature}
\bvolume{578}(\bissue{7793}),
\bfpage{60}--\blpage{65}
(\byear{2020}).
\doiurl{10.1038/s41586-020-1959-8}.
\end{barticle}
\endbibitem

\bibitem{king_algorithmic_2021}
\begin{barticle}
\bauthor{\bsnm{King}, \binits{S.A.}},
\bauthor{\bsnm{Spie\ss{}}, \binits{L.J.}},
\bauthor{\bsnm{Micke}, \binits{P.}},
\bauthor{\bsnm{Wilzewski}, \binits{A.}},
\bauthor{\bsnm{Leopold}, \binits{T.}},
\bauthor{\bsnm{Crespo~L\'opez-Urrutia}, \binits{J.R.}},
\bauthor{\bsnm{Schmidt}, \binits{P.O.}}:
\batitle{Algorithmic {Ground}-{State} {Cooling} of {Weakly} {Coupled}
  {Oscillators} {Using} {Quantum} {Logic}}.
\bjtitle{Phys. Rev. X}
\bvolume{11},
\bfpage{041049}
(\byear{2021}).
\doiurl{10.1103/PhysRevX.11.041049}
\end{barticle}
\endbibitem

\bibitem{lange_improved_2021}
\begin{barticle}
\bauthor{\bsnm{Lange}, \binits{R.}},
\bauthor{\bsnm{Huntemann}, \binits{N.}},
\bauthor{\bsnm{Rahm}, \binits{J.M.}},
\bauthor{\bsnm{Sanner}, \binits{C.}},
\bauthor{\bsnm{Shao}, \binits{H.}},
\bauthor{\bsnm{Lipphardt}, \binits{B.}},
\bauthor{\bsnm{Tamm}, \binits{Chr.}},
\bauthor{\bsnm{Weyers}, \binits{S.}},
\bauthor{\bsnm{Peik}, \binits{E.}}:
\batitle{Improved {Limits} for {Violations} of {Local} {Position} {Invariance}
  from {Atomic} {Clock} {Comparisons}}.
\bjtitle{Phys. Rev. Lett.}
\bvolume{126}(\bissue{1}),
\bfpage{011102}
(\byear{2021}).
\doiurl{10.1103/PhysRevLett.126.011102}.
\end{barticle}
\endbibitem

\bibitem{agababaev_ground-state_2018}
\begin{barticle}
\bauthor{\bsnm{Agababaev}, \binits{V.A.}},
\bauthor{\bsnm{Glazov}, \binits{D.A.}},
\bauthor{\bsnm{Volotka}, \binits{A.V.}},
\bauthor{\bsnm{Zinenko}, \binits{D.V.}},
\bauthor{\bsnm{Shabaev}, \binits{V.M.}},
\bauthor{\bsnm{Plunien}, \binits{G.}}:
\batitle{Ground-state $g$ factor of middle-{Z} boronlike ions}.
\bjtitle{J. Phys.: Conf. Ser. }
\bvolume{1138},
\bfpage{012003}
(\byear{2018}).
\doiurl{10.1088/1742-6596/1138/1/012003}
\end{barticle}
\endbibitem

\bibitem{agababaev_g_2019}
\begin{botherref}
\oauthor{\bsnm{Agababaev}, \binits{V.A.}},
\oauthor{\bsnm{Glazov}, \binits{D.A.}},
\oauthor{\bsnm{Volotka}, \binits{A.V.}},
\oauthor{\bsnm{Zinenko}, \binits{D.V.}},
\oauthor{\bsnm{Shabaev}, \binits{V.M.}},
\oauthor{\bsnm{Plunien}, \binits{G.}}:
$g$ factor of the
  [(1s)$^{\textrm{2}}$(2s)$^{\textrm{2}}$2p]$^{\textrm{2}}${P}$_{\textrm{3/2}}$
  state of middle-{Z} boronlike ions.
X-Ray Spectrometry
\textbf{1-6}(0)
(2019).
\doiurl{10.1002/xrs.3074}.
\end{botherref}
\endbibitem

\bibitem{yu_investigating_2019}
\begin{barticle}
\bauthor{\bsnm{Yu}, \binits{Y.-m.}},
\bauthor{\bsnm{Sahoo}, \binits{B.K.}}:
\batitle{Investigating ground-state fine-structure properties to explore
  suitability of boronlike {S}$^{11+}$-{K}$^{14+}$ and galliumlike
  {N}b$^{10+}$-{R}u$^{13+}$ ions as possible atomic clocks}.
\bjtitle{Phys. Rev. A}
\bvolume{99},
\bfpage{022513}
(\byear{2019}).
\doiurl{10.1103/PhysRevA.99.022513}.
\bcomment{(Note: units for quadrupole moment in Tab. VI. should read $\times
  10^{-7}$~Hz/(V/m$^{2}$) (Y-M. Yu, priv. comm.))}
\end{barticle}
\endbibitem

\bibitem{robert_priv_2021}
\begin{botherref}
\oauthor{\bsnm{M\"uller}, \binits{R.A.}}:
priv. comm.
(2021)
\end{botherref}
\endbibitem

\bibitem{micke_heidelberg_2018}
\begin{barticle}
\bauthor{\bsnm{Micke}, \binits{P.}},
\bauthor{\bsnm{Kühn}, \binits{S.}},
\bauthor{\bsnm{Buchauer}, \binits{L.}},
\bauthor{\bsnm{Harries}, \binits{J.R.}},
\bauthor{\bsnm{Bücking}, \binits{T.M.}},
\bauthor{\bsnm{Blaum}, \binits{K.}},
\bauthor{\bsnm{Cieluch}, \binits{A.}},
\bauthor{\bsnm{Egl}, \binits{A.}},
\bauthor{\bsnm{Hollain}, \binits{D.}},
\bauthor{\bsnm{Kraemer}, \binits{S.}},
\bauthor{\bsnm{Pfeifer}, \binits{T.}},
\bauthor{\bsnm{Schmidt}, \binits{P.O.}},
\bauthor{\bsnm{Schüssler}, \binits{R.X.}},
\bauthor{\bsnm{Schweiger}, \binits{C.}},
\bauthor{\bsnm{Stöhlker}, \binits{T.}},
\bauthor{\bsnm{Sturm}, \binits{S.}},
\bauthor{\bsnm{Wolf}, \binits{R.N.}},
\bauthor{\bsnm{Bernitt}, \binits{S.}},
\bauthor{\bsnm{Crespo~López-Urrutia}, \binits{J.R.}}:
\batitle{The {Heidelberg} compact electron beam ion traps}.
\bjtitle{Rev. Sci. Instrum}
\bvolume{89}(\bissue{6}),
\bfpage{063109}
(\byear{2018}).
\doiurl{10.1063/1.5026961}.
\end{barticle}
\endbibitem

\bibitem{leopold_cryogenic_2019}
\begin{barticle}
\bauthor{\bsnm{Leopold}, \binits{T.}},
\bauthor{\bsnm{King}, \binits{S.A.}},
\bauthor{\bsnm{Micke}, \binits{P.}},
\bauthor{\bsnm{Bautista-Salvador}, \binits{A.}},
\bauthor{\bsnm{Heip}, \binits{J.C.}},
\bauthor{\bsnm{Ospelkaus}, \binits{C.}},
\bauthor{\bsnm{Crespo~López-Urrutia}, \binits{J.R.}},
\bauthor{\bsnm{Schmidt}, \binits{P.O.}}:
\batitle{A cryogenic radio-frequency ion trap for quantum logic spectroscopy of
  highly charged ions}.
\bjtitle{Rev. Sci. Instrum}
\bvolume{90}(\bissue{7}),
\bfpage{073201}
(\byear{2019}).
\doiurl{10.1063/1.5100594}.
\end{barticle}
\endbibitem

\bibitem{micke_closed-cycle_2019}
\begin{barticle}
\bauthor{\bsnm{Micke}, \binits{P.}},
\bauthor{\bsnm{Stark}, \binits{J.}},
\bauthor{\bsnm{King}, \binits{S.A.}},
\bauthor{\bsnm{Leopold}, \binits{T.}},
\bauthor{\bsnm{Pfeifer}, \binits{T.}},
\bauthor{\bsnm{Schmöger}, \binits{L.}},
\bauthor{\bsnm{Schwarz}, \binits{M.}},
\bauthor{\bsnm{Spieß}, \binits{L.J.}},
\bauthor{\bsnm{Schmidt}, \binits{P.O.}},
\bauthor{\bsnm{Crespo~López-Urrutia}, \binits{J.R.}}:
\batitle{Closed-cycle, low-vibration 4 {K} cryostat for ion traps and other
  applications}.
\bjtitle{Rev. Sci. Instrum}
\bvolume{90}(\bissue{6}),
\bfpage{065104}
(\byear{2019}).
\doiurl{10.1063/1.5088593}.
\end{barticle}
\endbibitem

\bibitem{schmidt_spectroscopy_2005}
\begin{barticle}
\bauthor{\bsnm{Schmidt}, \binits{P.O.}},
\bauthor{\bsnm{Rosenband}, \binits{T.}},
\bauthor{\bsnm{Langer}, \binits{C.}},
\bauthor{\bsnm{Itano}, \binits{W.M.}},
\bauthor{\bsnm{Bergquist}, \binits{J.C.}},
\bauthor{\bsnm{Wineland}, \binits{D.J.}}:
\batitle{Spectroscopy {Using} {Quantum} {Logic}}.
\bjtitle{Science}
\bvolume{309}(\bissue{5735}),
\bfpage{749}--\blpage{752}
(\byear{2005}).
\doiurl{10.1126/science.1114375}.
\end{barticle}
\endbibitem

\bibitem{matei_1.5_2017}
\begin{barticle}
\bauthor{\bsnm{Matei}, \binits{D.G.}},
\bauthor{\bsnm{Legero}, \binits{T.}},
\bauthor{\bsnm{Häfner}, \binits{S.}},
\bauthor{\bsnm{Grebing}, \binits{C.}},
\bauthor{\bsnm{Weyrich}, \binits{R.}},
\bauthor{\bsnm{Zhang}, \binits{W.}},
\bauthor{\bsnm{Sonderhouse}, \binits{L.}},
\bauthor{\bsnm{Robinson}, \binits{J.M.}},
\bauthor{\bsnm{Ye}, \binits{J.}},
\bauthor{\bsnm{Riehle}, \binits{F.}},
\bauthor{\bsnm{Sterr}, \binits{U.}}:
\batitle{1.5 μm {Lasers} with {Sub}-10 {mHz} {Linewidth}}.
\bjtitle{Phys. Rev. Lett.}
\bvolume{118}(\bissue{26}),
\bfpage{263202}
(\byear{2017}).
\doiurl{10.1103/PhysRevLett.118.263202}.
\end{barticle}
\endbibitem

\bibitem{benkler_end--end_2019}
\begin{barticle}
\bauthor{\bsnm{Benkler}, \binits{E.}},
\bauthor{\bsnm{Lipphardt}, \binits{B.}},
\bauthor{\bsnm{Puppe}, \binits{T.}},
\bauthor{\bsnm{Wilk}, \binits{R.}},
\bauthor{\bsnm{Rohde}, \binits{F.}},
\bauthor{\bsnm{Sterr}, \binits{U.}}:
\batitle{End-to-end topology for fiber comb based optical frequency transfer at
  the $10^{-21}$ level}.
\bjtitle{Opt. Express}
\bvolume{27}(\bissue{25}),
\bfpage{36886}--\blpage{36902}
(\byear{2019}).
\doiurl{10.1364/OE.27.036886}
\end{barticle}
\endbibitem

\bibitem{lapierre_relativistic_2005}
\begin{botherref}
\oauthor{\bsnm{Lapierre}, \binits{A.}},
\oauthor{\bsnm{Jentschura}, \binits{U.}},
\oauthor{\bsnm{Crespo~López-Urrutia}, \binits{J.}},
\oauthor{\bsnm{Braun}, \binits{J.}},
\oauthor{\bsnm{Brenner}, \binits{G.}},
\oauthor{\bsnm{Bruhns}, \binits{H.}},
\oauthor{\bsnm{Fischer}, \binits{D.}},
\oauthor{\bsnm{González~Martínez}, \binits{A.}},
\oauthor{\bsnm{Harman}, \binits{Z.}},
\oauthor{\bsnm{Johnson}, \binits{W.}},
\oauthor{\bsnm{Keitel}, \binits{C.}},
\oauthor{\bsnm{Mironov}, \binits{V.}},
\oauthor{\bsnm{Osborne}, \binits{C.}},
\oauthor{\bsnm{Sikler}, \binits{G.}},
\oauthor{\bsnm{Soria~Orts}, \binits{R.}},
\oauthor{\bsnm{Shabaev}, \binits{V.}},
\oauthor{\bsnm{Tawara}, \binits{H.}},
\oauthor{\bsnm{Tupitsyn}, \binits{I.}},
\oauthor{\bsnm{Ullrich}, \binits{J.}},
\oauthor{\bsnm{Volotka}, \binits{A.}}:
Relativistic {Electron} {Correlation}, {Quantum} {Electrodynamics}, and the
  {Lifetime} of the 1s$^2$2s$^2$2p $^2${$P$}$^o_{3/2}$ {Level} in {Boronlike} {Argon}.
\bjtitle{Phys. Rev. Lett.}
\textbf{95}(18)
(2005).
\doiurl{10.1103/PhysRevLett.95.183001}
\end{botherref}
\endbibitem

\bibitem{peik_laser_2006}
\begin{barticle}
\bauthor{\bsnm{Peik}, \binits{E.}},
\bauthor{\bsnm{Schneider}, \binits{T.}},
\bauthor{\bsnm{Tamm}, \binits{Chr.}}:
\batitle{Laser frequency stabilization to a single ion}.
\bjtitle{J. Phys. B: At. Mol. Opt. Phys.}
\bvolume{39}(\bissue{1}),
\bfpage{145}--\blpage{158}
(\byear{2006}).
\doiurl{10.1088/0953-4075/39/1/012}.
\end{barticle}
\endbibitem

\bibitem{rosenband_frequency_2008}
\begin{barticle}
\bauthor{\bsnm{Rosenband}, \binits{T.}},
\bauthor{\bsnm{Hume}, \binits{D.B.}},
\bauthor{\bsnm{Schmidt}, \binits{P.O.}},
\bauthor{\bsnm{Chou}, \binits{C.W.}},
\bauthor{\bsnm{Brusch}, \binits{A.}},
\bauthor{\bsnm{Lorini}, \binits{L.}},
\bauthor{\bsnm{Oskay}, \binits{W.H.}},
\bauthor{\bsnm{Drullinger}, \binits{R.E.}},
\bauthor{\bsnm{Fortier}, \binits{T.M.}},
\bauthor{\bsnm{Stalnaker}, \binits{J.E.}},
\bauthor{\bsnm{Diddams}, \binits{S.A.}},
\bauthor{\bsnm{Swann}, \binits{W.C.}},
\bauthor{\bsnm{Newbury}, \binits{N.R.}},
\bauthor{\bsnm{Itano}, \binits{W.M.}},
\bauthor{\bsnm{Wineland}, \binits{D.J.}},
\bauthor{\bsnm{Bergquist}, \binits{J.C.}}:
\batitle{Frequency {Ratio} of {Al}$^{\textrm{+}}$ and {Hg}$^{\textrm{+}}$
  {Single}-{Ion} {Optical} {Clocks}; {Metrology} at the 17$^{\textrm{th}}$
  {Decimal} {Place}}.
\bjtitle{Science}
\bvolume{319}(\bissue{5871}),
\bfpage{1808}--\blpage{1812}
(\byear{2008}).
\doiurl{10.1126/science.1154622}
\end{barticle}
\endbibitem

\bibitem{dube_evaluation_2013}
\begin{barticle}
\bauthor{\bsnm{Dubé}, \binits{P.}},
\bauthor{\bsnm{Madej}, \binits{A.A.}},
\bauthor{\bsnm{Zhou}, \binits{Z.}},
\bauthor{\bsnm{Bernard}, \binits{J.E.}}:
\batitle{Evaluation of systematic shifts of the
  $^{\textrm{88}}${Sr}$^{\textrm{+}}$ single-ion optical frequency standard at
  the 10$^{\textrm{-17}}$ level}.
\bjtitle{Phys. Rev. A}
\bvolume{87}(\bissue{2}),
\bfpage{023806}
(\byear{2013}).
\doiurl{10.1103/PhysRevA.87.023806}.
\end{barticle}
\endbibitem

\bibitem{keller_precise_2015}
\begin{barticle}
\bauthor{\bsnm{Keller}, \binits{J.}},
\bauthor{\bsnm{Partner}, \binits{H.L.}},
\bauthor{\bsnm{Burgermeister}, \binits{T.}},
\bauthor{\bsnm{Mehlstäubler}, \binits{T.E.}}:
\batitle{Precise determination of micromotion for trapped-ion optical clocks}.
\bjtitle{J. Appl. Phys.}
\bvolume{118}(\bissue{10}),
\bfpage{104501}
(\byear{2015}).
\doiurl{10.1063/1.4930037}.
\end{barticle}
\endbibitem

\bibitem{beloy_frequency_2021}
\begin{barticle}
\bauthor{\bsnm{Beloy}, \binits{K.}},
\bauthor{\bsnm{Bodine}, \binits{M.I.}},
\bauthor{\bsnm{Bothwell}, \binits{T.}},
\bauthor{\bsnm{Brewer}, \binits{S.M.}},
\bauthor{\bsnm{Bromley}, \binits{S.L.}},
\bauthor{\bsnm{Chen}, \binits{J.-S.}},
\bauthor{\bsnm{Desch{\^e}nes}, \binits{J.-D.}},
\bauthor{\bsnm{Diddams}, \binits{S.A.}},
\bauthor{\bsnm{Fasano}, \binits{R.J.}},
\bauthor{\bsnm{Fortier}, \binits{T.M.}},
\bauthor{\bsnm{Hassan}, \binits{Y.S.}},
\bauthor{\bsnm{Hume}, \binits{D.B.}},
\bauthor{\bsnm{Kedar}, \binits{D.}},
\bauthor{\bsnm{Kennedy}, \binits{C.J.}},
\bauthor{\bsnm{Khader}, \binits{I.}},
\bauthor{\bsnm{Koepke}, \binits{A.}},
\bauthor{\bsnm{Leibrandt}, \binits{D.R.}},
\bauthor{\bsnm{Leopardi}, \binits{H.}},
\bauthor{\bsnm{Ludlow}, \binits{A.D.}},
\bauthor{\bsnm{McGrew}, \binits{W.F.}},
\bauthor{\bsnm{Milner}, \binits{W.R.}},
\bauthor{\bsnm{Newbury}, \binits{N.R.}},
\bauthor{\bsnm{Nicolodi}, \binits{D.}},
\bauthor{\bsnm{Oelker}, \binits{E.}},
\bauthor{\bsnm{Parker}, \binits{T.E.}},
\bauthor{\bsnm{Robinson}, \binits{J.M.}},
\bauthor{\bsnm{Romisch}, \binits{S.}},
\bauthor{\bsnm{Sch{\"a}ffer}, \binits{S.A.}},
\bauthor{\bsnm{Sherman}, \binits{J.A.}},
\bauthor{\bsnm{Sinclair}, \binits{L.C.}},
\bauthor{\bsnm{Sonderhouse}, \binits{L.}},
\bauthor{\bsnm{Swann}, \binits{W.C.}},
\bauthor{\bsnm{Yao}, \binits{J.}},
\bauthor{\bsnm{Ye}, \binits{J.}},
\bauthor{\bsnm{Zhang}, \binits{X.}},
\bauthor{\bsnm{{Boulder Atomic Clock Optical Network (BACON) Collaboration*}}}:
\batitle{Frequency ratio measurements at 18-digit accuracy using an optical
  clock network}.
\bjtitle{Nature}
\bvolume{591}(\bissue{7851}),
\bfpage{564}--\blpage{569}
(\byear{2021}).
\doiurl{10.1038/s41586-021-03253-4}
\end{barticle}
\endbibitem

\bibitem{nemitz_absolute_2021}
\begin{barticle}
\bauthor{\bsnm{Nemitz}, \binits{N.}},
\bauthor{\bsnm{Gotoh}, \binits{T.}},
\bauthor{\bsnm{Nakagawa}, \binits{F.}},
\bauthor{\bsnm{Ito}, \binits{H.}},
\bauthor{\bsnm{Hanado}, \binits{Y.}},
\bauthor{\bsnm{Ido}, \binits{T.}},
\bauthor{\bsnm{Hachisu}, \binits{H.}}:
\batitle{Absolute {F}requency of {{$^{87}$Sr}} at 1.8 \texttimes{} 10$^{-16}$
  {U}ncertainty by {R}eference to {R}emote {P}rimary {F}requency {S}tandards}.
\bjtitle{Metrologia}
\bvolume{58}(\bissue{2}),
\bfpage{025006}
(\byear{2021}).
\doiurl{10.1088/1681-7575/abc232}
\end{barticle}
\endbibitem

\bibitem{pizzocaro_absolute_2020}
\begin{barticle}
\bauthor{\bsnm{Pizzocaro}, \binits{M.}},
\bauthor{\bsnm{Bregolin}, \binits{F.}},
\bauthor{\bsnm{Barbieri}, \binits{P.}},
\bauthor{\bsnm{Rauf}, \binits{B.}},
\bauthor{\bsnm{Levi}, \binits{F.}},
\bauthor{\bsnm{Calonico}, \binits{D.}}:
\batitle{Absolute {F}requency {M}easurement of the
  {{$^1$S$_0$}}\textendash{{$^3$P$_0$}} transition of {{$^{171}$Yb}} with a
  {L}ink to {I}nternational {A}tomic {T}ime}.
\bjtitle{Metrologia}
\bvolume{57}(\bissue{3}),
\bfpage{035007}
(\byear{2020}).
\doiurl{10.1088/1681-7575/ab50e8}
\end{barticle}
\endbibitem

\bibitem{shabaev_qed_1998}
\begin{barticle}
\bauthor{\bsnm{Shabaev}, \binits{V.M.}}:
\batitle{{QED} theory of the nuclear recoil effect in atoms}.
\bjtitle{Phys. Rev. A}
\bvolume{57},
\bfpage{59}--\blpage{67}
(\byear{1998}).
\doiurl{10.1103/PhysRevA.57.59}
\end{barticle}
\endbibitem

\bibitem{arapoglou_g-factor_2019}
\begin{barticle}
\bauthor{\bsnm{Arapoglou}, \binits{I.}},
\bauthor{\bsnm{Egl}, \binits{A.}},
\bauthor{\bsnm{Höcker}, \binits{M.}},
\bauthor{\bsnm{Sailer}, \binits{T.}},
\bauthor{\bsnm{Tu}, \binits{B.}},
\bauthor{\bsnm{Weigel}, \binits{A.}},
\bauthor{\bsnm{Wolf}, \binits{R.}},
\bauthor{\bsnm{Cakir}, \binits{H.}},
\bauthor{\bsnm{Yerokhin}, \binits{V.A.}},
\bauthor{\bsnm{Oreshkina}, \binits{N.S.}},
\bauthor{\bsnm{Agababaev}, \binits{V.A.}},
\bauthor{\bsnm{Volotka}, \binits{A.V.}},
\bauthor{\bsnm{Zinenko}, \binits{D.V.}},
\bauthor{\bsnm{Glazov}, \binits{D.A.}},
\bauthor{\bsnm{Harman}, \binits{Z.}},
\bauthor{\bsnm{Keitel}, \binits{C.H.}},
\bauthor{\bsnm{Sturm}, \binits{S.}},
\bauthor{\bsnm{Blaum}, \binits{K.}}:
\batitle{\textit{g}-factor of {Boronlike} {Argon}
  $^{\textrm{40}}${Ar}$^{\textrm{13+}}$}.
\bjtitle{Phys. Rev. Lett.}
\bvolume{122}(\bissue{25}),
\bfpage{253001}
(\byear{2019}).
\doiurl{10.1103/PhysRevLett.122.253001}.
\end{barticle}
\endbibitem

\bibitem{naze_isotope_2014}
\begin{barticle}
\bauthor{\bsnm{Nazé}, \binits{C.}},
\bauthor{\bsnm{Verdebout}, \binits{S.}},
\bauthor{\bsnm{Rynkun}, \binits{P.}},
\bauthor{\bsnm{Gaigalas}, \binits{G.}},
\bauthor{\bsnm{Godefroid}, \binits{M.}},
\bauthor{\bsnm{Jönsson}, \binits{P.}}:
\batitle{Isotope shifts in beryllium-, boron-, carbon-, and nitrogen-like ions
  from relativistic configuration interaction calculations}.
\bjtitle{Atomic Data and Nuclear Data Tables}
\bvolume{100}(\bissue{5}),
\bfpage{1197}--\blpage{1249}
(\byear{2014}).
\doiurl{10.1016/j.adt.2014.02.004}.
\end{barticle}
\endbibitem

\bibitem{yu_selected_2018}
\begin{barticle}
\bauthor{\bsnm{Yu}, \binits{Y.-m.}},
\bauthor{\bsnm{Sahoo}, \binits{B.K.}}:
\batitle{Selected highly charged ions as prospective candidates for optical
  clocks with quality factors larger than 10$^{\textrm{15}}$}.
\bjtitle{Phys. Rev. A}
\bvolume{97}(\bissue{4}),
\bfpage{041403}
(\byear{2018}).
\doiurl{10.1103/PhysRevA.97.041403}.
\end{barticle}
\endbibitem

\bibitem{bekker_detection_2019}
\begin{barticle}
\bauthor{\bsnm{Bekker}, \binits{H.}},
\bauthor{\bsnm{Borschevsky}, \binits{A.}},
\bauthor{\bsnm{Harman}, \binits{Z.}},
\bauthor{\bsnm{Keitel}, \binits{C.H.}},
\bauthor{\bsnm{Pfeifer}, \binits{T.}},
\bauthor{\bsnm{Schmidt}, \binits{P.O.}},
\bauthor{\bsnm{López-Urrutia}, \binits{J.R.C.}},
\bauthor{\bsnm{Berengut}, \binits{J.C.}}:
\batitle{Detection of the 5p – 4f orbital crossing and its optical clock
  transition in {Pr}$^{\textrm{9+}}$}.
\bjtitle{Nat. Commun}
\bvolume{10}(\bissue{1}),
\bfpage{5651}
(\byear{2019}).
\doiurl{10.1038/s41467-019-13406-9}.
\end{barticle}
\endbibitem

\bibitem{berengut_generalized_2020}
\begin{barticle}
\bauthor{\bsnm{Berengut}, \binits{J.C.}},
\bauthor{\bsnm{Delaunay}, \binits{C.}},
\bauthor{\bsnm{Geddes}, \binits{A.}},
\bauthor{\bsnm{Soreq}, \binits{Y.}}:
\batitle{{Generalized} {King} linearity and new physics searches with isotope
  shifts}.
\bjtitle{Phys. Rev. Research}
\bvolume{2},
\bfpage{043444}
(\byear{2020}).
\doiurl{10.1103/PhysRevResearch.2.043444}
\end{barticle}
\endbibitem

\bibitem{rehbehn_sensitivity_2021}
\begin{barticle}
\bauthor{\bsnm{Rehbehn}, \binits{N.-H.}},
\bauthor{\bsnm{Rosner}, \binits{M.K.}},
\bauthor{\bsnm{Bekker}, \binits{H.}},
\bauthor{\bsnm{Berengut}, \binits{J.C.}},
\bauthor{\bsnm{Schmidt}, \binits{P.O.}},
\bauthor{\bsnm{King}, \binits{S.A.}},
\bauthor{\bsnm{Micke}, \binits{P.}},
\bauthor{\bsnm{Gu}, \binits{M.F.}},
\bauthor{\bsnm{M{\"u}ller}, \binits{R.}},
\bauthor{\bsnm{Surzhykov}, \binits{A.}},
\bauthor{\bsnm{{L{\'o}pez-Urrutia}}, \binits{J.R.C.}}:
\batitle{Sensitivity to new physics of isotope-shift studies using the coronal
  lines of highly charged calcium ions}.
\bjtitle{Phys. Rev. A}
\bvolume{103}(\bissue{4}),
\bfpage{040801}
(\byear{2021}).
\doiurl{10.1103/PhysRevA.103.L040801}
\end{barticle}
\endbibitem

\bibitem{dzuba_highly_2015}
\begin{barticle}
\bauthor{\bsnm{Dzuba}, \binits{V.A.}},
\bauthor{\bsnm{Flambaum}, \binits{V.V.}}:
\batitle{Highly charged ions for atomic clocks and search for variation of the
  fine structure constant}.
\bjtitle{Hyperfine Interact}
\bvolume{236}(\bissue{1-3}),
\bfpage{79}--\blpage{86}
(\byear{2015}).
\doiurl{10.1007/s10751-015-1166-4}.
\end{barticle}
\endbibitem

\bibitem{porsev_optical_2020}
\begin{barticle}
\bauthor{\bsnm{Porsev}, \binits{S.G.}},
\bauthor{\bsnm{Safronova}, \binits{U.I.}},
\bauthor{\bsnm{Safronova}, \binits{M.S.}},
\bauthor{\bsnm{Schmidt}, \binits{P.O.}},
\bauthor{\bsnm{Bondarev}, \binits{A.I.}},
\bauthor{\bsnm{Kozlov}, \binits{M.G.}},
\bauthor{\bsnm{Tupitsyn}, \binits{I.I.}},
\bauthor{\bsnm{Cheung}, \binits{C.}}:
\batitle{Optical clocks based on the {Cf}$^{\textrm{15+}}$ and
  {Cf}$^{\textrm{17+}}$ ions}.
\bjtitle{Phys. Rev. A}
\bvolume{102}(\bissue{1}),
\bfpage{012802}
(\byear{2020}).
\doiurl{10.1103/PhysRevA.102.012802}.
\end{barticle}
\endbibitem

\bibitem{itano_external-field_2000}
\begin{barticle}
\bauthor{\bsnm{Itano}, \binits{W.M.}}:
\batitle{External-field shifts of the $^{\textrm{199}}${Hg}$^{\textrm{+}}$
  optical frequency standard}.
\bjtitle{J. Res. Natl. Inst. Stand. Technol.}
\bvolume{105}(\bissue{6}),
\bfpage{829}--\blpage{837}
(\byear{2000}).
\doiurl{10.6028/jres.105.065}
\end{barticle}
\endbibitem

\bibitem{akerman_atomic_2018}
\begin{barticle}
\bauthor{\bsnm{Akerman}, \binits{N.}},
\bauthor{\bsnm{Ozeri}, \binits{R.}}:
\batitle{Atomic combination clocks}.
\bjtitle{New J. Phys.}
\bvolume{20}(\bissue{12}),
\bfpage{123026}
(\byear{2018}).
\doiurl{10.1088/1367-2630/aaf4cb}.
\end{barticle}
\endbibitem

\bibitem{Gan2018}
\begin{barticle}
\bauthor{\bsnm{Gan}, \binits{H.C.J.}},
\bauthor{\bsnm{Maslennikov}, \binits{G.}},
\bauthor{\bsnm{Tseng}, \binits{K.-W.}},
\bauthor{\bsnm{Tan}, \binits{T.R.}},
\bauthor{\bsnm{Kaewuam}, \binits{R.}},
\bauthor{\bsnm{Arnold}, \binits{K.J.}},
\bauthor{\bsnm{Matsukevich}, \binits{D.}},
\bauthor{\bsnm{Barrett}, \binits{M.D.}}:
\batitle{Oscillating-magnetic-field effects in high-precision metrology}.
\bjtitle{Phys. Rev. A}
\bvolume{98},
\bfpage{032514}
(\byear{2018}).
\doiurl{10.1103/PhysRevA.98.032514}
\end{barticle}
\endbibitem

\bibitem{arnold_precision_2020}
\begin{barticle}
\bauthor{\bsnm{Arnold}, \binits{K.J.}},
\bauthor{\bsnm{Kaewuam}, \binits{R.}},
\bauthor{\bsnm{Chanu}, \binits{S.R.}},
\bauthor{\bsnm{Tan}, \binits{T.R.}},
\bauthor{\bsnm{Zhang}, \binits{Z.}},
\bauthor{\bsnm{Barrett}, \binits{M.D.}}:
\batitle{Precision {Measurements} of the {$^{138}{\mathrm{Ba}}^{+}$}
  {$6s{^{2}S}_{1/2}\ensuremath{-}5d{^{2}D}_{5/2}$} {Clock} {Transition}}.
\bjtitle{Phys. Rev. Lett.}
\bvolume{124},
\bfpage{193001}
(\byear{2020}).
\doiurl{10.1103/PhysRevLett.124.193001}
\end{barticle}
\endbibitem

\bibitem{yerokhin_nonlinear_2020}
\begin{barticle}
\bauthor{\bsnm{Yerokhin}, \binits{V.A.}},
\bauthor{\bsnm{Müller}, \binits{R.A.}},
\bauthor{\bsnm{Surzhykov}, \binits{A.}},
\bauthor{\bsnm{Micke}, \binits{P.}},
\bauthor{\bsnm{Schmidt}, \binits{P.O.}}:
\batitle{Nonlinear isotope-shift effects in {Be}-like, {B}-like, and {C}-like
  argon}.
\bjtitle{Phys. Rev. A}
\bvolume{101}(\bissue{1}),
\bfpage{012502}
(\byear{2020}).
\doiurl{10.1103/PhysRevA.101.012502}.
\end{barticle}
\endbibitem

\bibitem{shabaev_model_2013}
\begin{barticle}
\bauthor{\bsnm{Shabaev}, \binits{V.M.}},
\bauthor{\bsnm{Tupitsyn}, \binits{I.I.}},
\bauthor{\bsnm{Yerokhin}, \binits{V.A.}}:
\batitle{Model operator approach to the {L}amb shift calculations in
  relativistic many-electron atoms}.
\bjtitle{Phys. Rev. A}
\bvolume{88},
\bfpage{012513}
(\byear{2013}).
\doiurl{10.1103/PhysRevA.88.012513}
\end{barticle}
\endbibitem

\bibitem{shabaev_qedmod_2015}
\begin{barticle}
\bauthor{\bsnm{Shabaev}, \binits{V.M.}},
\bauthor{\bsnm{Tupitsyn}, \binits{I.I.}},
\bauthor{\bsnm{Yerokhin}, \binits{V.A.}}:
\batitle{{QEDMOD}: {F}ortran program for calculating the model {L}amb-shift
  operator}.
\bjtitle{Comput. Phys. Commun}
\bvolume{189},
\bfpage{175}--\blpage{181}
(\byear{2015}).
\doiurl{10.1016/j.cpc.2014.12.002}
\end{barticle}
\endbibitem

\bibitem{yerokhin_nuclear-size_2011}
\begin{botherref}
\oauthor{\bsnm{Yerokhin}, \binits{V.A.}}:
Nuclear-size correction to the {Lamb} shift of one-electron atoms.
\bjtitle{Phys. Rev. A}
\textbf{83}(1)
(2011).
\doiurl{10.1103/PhysRevA.83.012507}.
\end{botherref}
\endbibitem

\bibitem{angeli_table_2013}
\begin{barticle}
\bauthor{\bsnm{Angeli}, \binits{I.}},
\bauthor{\bsnm{Marinova}, \binits{K.P.}}:
\batitle{Table of experimental nuclear ground state charge radii: {An} update}.
\bjtitle{Atomic Data and Nuclear Data Tables}
\bvolume{99}(\bissue{1}),
\bfpage{69}--\blpage{95}
(\byear{2013}).
\doiurl{10.1016/j.adt.2011.12.006}.
\end{barticle}
\endbibitem

\bibitem{wang_ame2012_2012}
\begin{barticle}
\bauthor{\bsnm{Wang}, \binits{M.}},
\bauthor{\bsnm{Audi}, \binits{G.}},
\bauthor{\bsnm{Wapstra}, \binits{A.H.}},
\bauthor{\bsnm{Kondev}, \binits{F.G.}},
\bauthor{\bsnm{MacCormick}, \binits{M.}},
\bauthor{\bsnm{Xu}, \binits{X.}},
\bauthor{\bsnm{Pfeiffer}, \binits{B.}}:
\batitle{The {A}me2012 atomic mass evaluation}.
\bjtitle{Chin. Phys. C}
\bvolume{36}(\bissue{12}),
\bfpage{1603}--\blpage{2014}
(\byear{2012}).
\doiurl{10.1088/1674-1137/36/12/003}
\end{barticle}
\endbibitem

\end{thebibliography}
\end{document}

\title{Supplementary Material for An Optical Atomic Clock Based on a Highly Charged Ion} %

\author[1,5,$\dagger$]{\mbox{Steven A. King}}

\author[1,$\dagger$,*]{\mbox{Lukas J. Spieß}}

\author[1,2,6]{\mbox{Peter Micke}}

\author[1]{\mbox{Alexander Wilzewski}}

\author[1,7]{\mbox{Tobias Leopold}}

\author[1]{\mbox{Erik Benkler}}

\author[1]{\mbox{Richard Lange}}

\author[1]{\mbox{Nils Huntemann}}

\author[1,3]{\mbox{Andrey Surzhykov}}

\author[1]{\mbox{Vladimir A. Yerokhin}}

\author[2]{\mbox{Jos\'e R. Crespo L\'opez-Urrutia}}

\author[1,4,*]{\mbox{Piet O. Schmidt}}

\affil[1]{Physikalisch-Technische Bundesanstalt, Braunschweig, Germany}

\affil[2]{Max-Planck-Institut f\"ur Kernphysik, Heidelberg, Germany}

\affil[3]{Technische Universit\"at Braunschweig, Braunschweig, Germany}

\affil[4]{Institut f\"ur Quantenoptik, Hannover, Germany}

\affil[5]{Present Address: Oxford Ionics, Begbroke, United Kingdom}
\affil[6]{Present Address: CERN, Geneva, Switzerland}
\affil[7]{\vspace{2ex} Present Address: LPKF Laser \& Electronics AG, Garbsen, Germany}

\affil[*]{Corresponding authors. Email: lukas.spiess@quantummetrology.de, piet.schmidt@quantummetrology.de}

\affil[$\dagger$]{These authors contributed equally to this work.}

\date{\vspace{-5ex}}

\maketitle

\textbf{In this supplementary document, we provide a comprehensive analysis of the systematic perturbations that must be taken into account as part of the frequency measurement process, along with further information about the theoretical calculation of the isotope shift. For highly charged ions, many of the stray-field-induced systematic shifts that affect singly-charged and neutral atoms are greatly suppressed as a consequence of their strong inverse scalings with the charge state of the atom \cite{berengut_highly_2012}}.

\section{Stark shifts}
Electric fields perturb the frequency of the clock transition through the a.c. Stark effect:
\begin{equation}
\Delta \nu_{\textrm{ACS}} = -\frac{1}{2h} \Delta \alpha \left< E^2 \right>
\end{equation}
where $\Delta \alpha$ is the difference between the polarisabilities of the upper and lower states involved in the clock transition, $\left< E^2 \right>$ is the mean-square amplitude of the perturbing field, $h$ is the Planck constant, and an unpolarised field has been assumed. The value of $\Delta \alpha$ depends on the frequency of the perturbing field, and tends to a steady-state value referred to as the static differential polarisability for large detunings below any atomic resonances.\par
In general, the value of the static differential polarisability is expected to be very small for optical clock transitions in highly charged ions, as the energies of $E1$-allowed transitions scale up with the charge state, whilst their dipole matrix elements are reduced. This results in a net scaling of $Z_a^{-4}$ for the differential polarisability, where $Z_a$ is the effective screened charge of the ion \cite{berengut_highly_2012}.
For \Ar, the lowest-frequency $E1$-allowed transitions for the ground and excited states of the clock transition have wavelengths of 19~nm and 48~nm respectively. Therefore the two states have small and very similar polarisabilities for the typical radio- and optical-frequency perturbing fields in ion trap experiments. This leads to a differential static polarisability of $-3\times10^{-45}$~Jm$^{2}$V$^{-2}$, or $-2\times10^{-4}$ in atomic units \cite{yu_investigating_2019}. This is two orders of magnitude smaller than the \oneszero{} $\rightarrow$ \threedone{} transition in \Lu{} \cite{arnold_blackbody_2018}, which has the smallest known differential polarisability amongst optical clocks based on singly-charged ions and neutral atoms. This practically eliminates Stark shifts from several common sources of electric fields, such as leaked laser radiation and residual trapping fields at the position of the HCI. A laser beam with a power of \SI{1}{\micro\watt} and a $1/e^2$ radius of \SI{30}{\micro\metre} would lead to a shift of only $1\times10^{-21}$, and a displacement of the ion from the rf null where it experiences an rf electric field of 100~V/m would lead to a shift of only $4\times10^{-23}$.

\subsection{Blackbody radiation shift}
\label{subsec:bbr}
Blackbody radiation (BBR) emitted by surfaces at finite temperatures couples to electric-dipole ($E1$) and magnetic-dipole ($M1$) transitions from the two electronic states involved in the clock transition, leading to a systematic shift of the two levels via the a.c. Stark and a.c. Zeeman effects, respectively.\par
The electric field of blackbody radiation is described by:
\begin{equation}
\left< E_{\textrm{BBR}}^2 \right> = (831.9\,\textrm{V/m})^2 \left( \frac{T}{300 \textrm{K}} \right)^4.
\label{eq:bbr}
\end{equation}
If every surface visible to the HCI was held at room temperature, this would lead to a BBR shift of only $2.5 \times 10^{-21}$. In actuality, the surfaces spanning the overwhelming majority of the solid angle visible to the ion are held at a temperature near 4~K, with room temperature surfaces comprising only approximately $10^{-3}$ of the $4\pi$ solid angle. This reduces the magnitude of the BBR shift from room-temperature radiation by a further three orders of magnitude. Emission from the cold surfaces is suppressed compared to room temperature by more than seven orders of magnitude according to Eq.~\ref{eq:bbr} and so can be neglected.\par %
The contribution from the magnetic field of the BBR was also calculated in Ref.~\cite{yu_investigating_2019} to be $-1.1 \times 10^{-21}$ at a temperature of 300~K. Similarly, this will be suppressed by around three orders of magnitude in our setup.

\subsection{Probe-laser-induced shifts}
The a.c. Stark effect from the probe laser can be neglected as described above. However, close to resonance, the laser will perturb the clock transition directly via a resonant a.c. Zeeman effect. When interrogating an individual Zeeman component of the transition at its two half-maxima points, the net resonant effect of the laser on the determined centre frequency is zero. However, the laser also couples off-resonantly to other nearby Zeeman components, leading to a power-dependent shift. 
The polarisation composition of the laser requires the laser power to be varied between Zeeman components in order to achieve an identical Rabi frequency, which is the condition for optimal measurement stability when averaging across several lines. This leads to a nonzero net shift on the mean frequency of the Zeeman components. Before the measurement, the polarisations of the two probe beams were optimised to give similar coupling strengths for each of the four Zeeman components used in this work.\par

In principle, this residual shift can be removed in real-time using more advanced interrogation schemes \cite{king_absolute_2012, huntemann_high-accuracy_2012, yudin_hyper-ramsey_2010, sanner_autobalanced_2018}. As the induced shift is small compared to the total systematic uncertainty, we pre-calibrated the effect through a zero-power extrapolation of shifts measured at high power. The measured perturbed mean transition frequency $f'$ at a probe power level $i$ can be described by
\begin{equation}
    f'_i=f_{0} + k_{\textrm{LS}} \Omega_{\textrm{Rabi},i}^2,
\end{equation}
where $f_0$ is the unperturbed frequency at zero laser power, $k_{\textrm{LS}}$ is a constant describing coupling to other states, and $\Omega_{\textrm{Rabi},i}^2$ as the square of the resonant Rabi frequency is directly proportional to the used laser power.\par

\begin{figure}[t]
    \begin{minipage}[t]{.01\textwidth}
        \textbf{a}
	\end{minipage}
    \begin{minipage}[t]{.49\textwidth}
        \centering \raisebox{\dimexpr-\height+1.5ex\relax}{\includegraphics[width=0.95\textwidth]{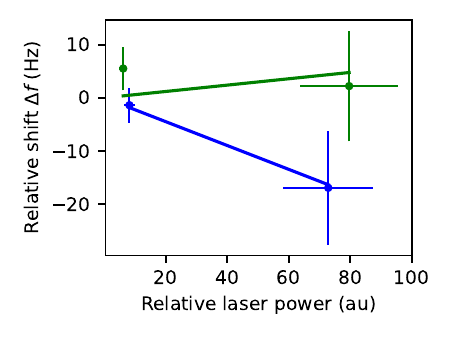}}
	\end{minipage}
    \begin{minipage}[t]{.01\textwidth}
        \textbf{b}
	\end{minipage}
    \begin{minipage}[t]{.49\textwidth}
        \centering \raisebox{\dimexpr-\height+1.5ex\relax}{\includegraphics[width=0.95\textwidth]{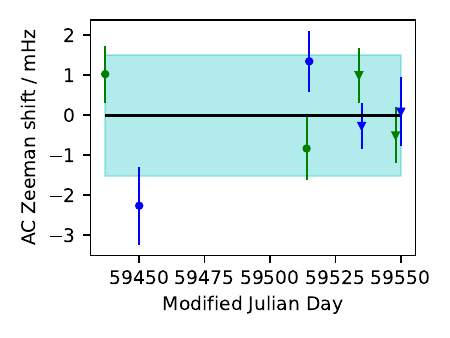}}
	\end{minipage}
	\caption{\textbf{Determination of the probe-laser-induced a.c. Zeeman shift.} The shift is estimated by measuring at various laser powers corresponding to different $\pi$-times $t_{\pi,i}$. The relative shift of these measurements is extrapolated to the much lower laser powers used during clock operation. Further details are given in the text. \textbf{a} Example measurement for each of the two counter-propagating probe beams (green/blue). The gradient is the only degree of freedom of the fit. \textbf{b} Extrapolated shifts at the 15~ms probe time for the two probe directions and for the two isotopes (\Arforty: circles, \Arthirtysix: triangles). The estimated 68\% confidence interval is indicated by the shaded background.}
	\label{fig:aczeeman}
\end{figure}

The constant $k_{\textrm{LS}}$ for a given probe direction was experimentally determined by running a modified version of the clock laser sequence. For each of the four Zeeman components, three servos were operated using levels of laser intensity that spanned several orders of magnitude. To eliminate the unknown $f_0$, the perturbed transition frequencies at the intermediate and high power levels ($f'_2$, $f'_3$) were compared to the value at the lowest laser power ($f'_1$) on a cycle-by-cycle basis. This frequency difference is linear in the difference between the two squared Rabi frequencies:
\begin{equation}
\Delta f_2 = f'_2 - f'_1 = k_{\textrm{LS}} (\Omega_2^2-\Omega_1^2) = k_{\textrm{LS}} \left( \frac{\pi}{(t_{\pi,2})^2}-\frac{\pi}{(t_{\pi,1})^2} \right),
\end{equation}
where $\Omega_{\textrm{Rabi},i}$ is replaced by the experimentally accessible $\pi$-time $t_\pi$. A similar expression exists for $\Delta f_3$. $k_{\textrm{LS}}$ can then be extracted from a linear fit to the data for the two frequency differences $\Delta f_2$ and $\Delta f_3$, as shown in Fig.~\ref{fig:aczeeman}a.

The measurement was performed before and after the absolute frequency measurements of each isotope and independently for each probe direction. The results are given in Fig.~\ref{fig:aczeeman}b. Typically a residual shift of $\pm1.5$~mHz for the two counterpropagating clock laser beams was determined after optimisation, corresponding to a fractional shift of $2.2\times10^{-18}$. Over timescales of a few weeks, changes on the order of a few mHz on this compensation were observed. This may have been caused by alteration of the birefringence in the laser windows on the cryogenic heatshield due to changes in mounting stress %
or laboratory temperature-induced changes in the waveplates used to set the polarisations of the two beams. The uncertainty was estimated based on the observed variations. %

\section{Magnetic-field-induced shifts}

\subsection{Magnetic field instability}
\label{subsec:bfield}
As the individual components of the \Ar{} clock transition are sensitive to magnetic fields in first order, the applied magnetic field used to define the quantisation axis must be held as stable as possible to avoid line-broadening effects or offsets on the clock laser frequency after locking. Our scheme for producing a highly stable bias field is described in Ref.~\cite{leopold_cryogenic_2019}. Since this publication, we have further improved the stability on the field by using a system similar to that presented in Ref.~\cite{merkel_magnetic_2019} to reduce the noise on the employed benchtop current supplies. The bandwidth of the feedback is also sufficient to suppress noise at the mains frequency of 50~Hz and its harmonics coming from the current supplies. In this manner, we are able to achieve a short-term fractional stability of around $10^{-7}$ on the individual currents, around an order of magnitude better than the free-running supply, with greatly improved mid- and long-term performance owing to the several orders of magnitude lower sensitivity of the sense resistor to the temperature of the laboratory compared to the benchtop power supply. We typically observe a field stability at the position of the ion that is better than 200~pT at averaging times between 1 and 1000~s, close to the limit imposed by the fluxgate magnetometer used to stabilise the field \cite{leopold_cryogenic_2019}.

\subsubsection{Linear Zeeman shift}
Both of the clock states in Ar$^{13+}$ have half-integer total electronic angular momentum. This leads to a first-order sensitivity to the magnetic flux density $B$:
\begin{equation}
	\Delta f^{(1)}_\text{Zeeman} = \frac{m_{J} \, g_{J} \, \mu_\text{B}}{h}  \, B\, ,
	\label{eq:linZeeman}
\end{equation}
where $m_{J}$ is the magnetic quantum number and $g_{J}$ is the Land\'e $g$-factor of the state, $\mu_{B}$ is the Bohr magneton, and $h$ is the Planck constant. The experiment employs a flux density of \SI{23}{\micro\tesla} to define the quantisation axis and to lift the degeneracy of the magnetic sublevels. This leads to a splitting of around 1~MHz between the outermost components of the clock transition, making it by far the largest systematic perturbation of the individual transition frequencies. Despite the relatively large magnitude of this shift, it can be suppressed by averaging over symmetric pairs of Zeeman components \cite{dube_evaluation_2013,huang_evaluation_2014}. The employed pseudo-randomisation of the order in which the Zeeman components are interrogated suppresses any drifts in the magnetic field between measurements.

\subsubsection{Quadratic Zeeman shift}

The magnetic quantisation field also shifts the clock transition via a second-order Zeeman effect. Here, the magnetic field shifts the involved states as \cite{itano_external-field_2000}:
\begin{equation}
	\frac{\Delta f^{\left( 2 \right)}_\text{Zeeman}}{f_0} = k_{\textrm{QZ}}(m_J) B^2,
	\label{eq:secZeeman}
\end{equation}
where the scaling constant $k_{\textrm{QZ}}$ can be calculated in perturbation theory for HCI with an uncertainty level of about $1\,\%$ \cite{lindenfels_experimental_2015}. Similarly to the general lack of $E1$-allowed optical transitions in \Ar{}, apart from the clock transition itself the lowest-frequency magnetic-dipole ($M1$)-allowed transitions from either of the two states involved have energies in the XUV range. This leads to values of $k_{\textrm{QZ}}$ that are on the order of \SI{1e-22}{\per\micro\tesla\squared} \cite{yu_investigating_2019,von_lindenfels_experimental_2013}, more than an order of magnitude smaller than for the \oneszero{} $\rightarrow$ \threepzero{} transition in \Hg{} \cite{hachisu_trapping_2008}, which has the smallest coefficient of any currently operational optical clock. This results in a shift below the $10^{-19}$ level at our operating field of \SI{23}{\micro\tesla}.\par

\subsection{Radiofrequency magnetic fields}
A quadratic Zeeman shift can also be induced by oscillating rf magnetic fields due to imbalanced currents in the rf electrodes of the trap \cite{Gan2018}. The magnitude of this field is difficult to determine in our system, but its component directed along the quantisation axis was previously measured to have a mean-square amplitude of less than \SI{3}{\micro\tesla\squared} when approximately twice the rf voltage was applied to the electrodes compared to that used in this work \cite{leopold_cryogenic_2019}. Due to the geometry of the trap, we assume the transverse components of the magnetic field to be of the same magnitude, leading to a total shift at the $10^{-21}$ level.\par

\section{Electric quadrupole shift}
HCI are predicted to possess small values for the electric quadrupole moment $\Theta$ in general, as a consequence of its $Z_a^{-2}$ scaling \cite{berengut_highly_2012}. Indeed, for \Ar{}, $\Theta = 0.0235\,ea_{0}^{2}$ is predicted for the excited \pthreehalves{} state \cite{yu_investigating_2019, robert_priv_2021}. This is slightly smaller than that of the \fsevenhalves{} state in \Yb{} \cite{lange_coherent_2020}, which is the smallest known amongst ion clocks that utilise states with nonzero quadrupole moments. \par
Based on this value of $\Theta$, we expect the magnitude of the shift to be approximately $10\,$mHz under our typical trapping conditions with $\partial E_z/\partial z \approx 2\times10^{6}$~Vm$^{-2}$. The shift was therefore not resolved at our statistically-limited uncertainty of approximately 100~mHz. Further, it is suppressed by probing each Zeeman substate of the excited state once, with the mean energy of the substates being free of the quadrupole shift. Similarly, to the linear Zeeman shift, any temporal variations are suppressed by pseudorandomisation. Assuming a suppression by at least a factor of 20 \cite{dube_evaluation_2013}, the residual shift is below $10^{-18}$ and therefore negligible.

\section{Motional shifts}
\subsection{First-order Doppler shift}
\label{subsec:firstdopp}
A first-order Doppler shift can arise from slow drifts of the ion position relative to the laboratory frame \cite{rosenband_frequency_2008}, which can be induced by slow charging of surfaces near to the ion or slight movements of the suspended cold stages inside the cryostat. In order to eliminate this, two counterpropagating directions are used for the probe beam, each of which has independent servos to track the frequency of the Zeeman components of the transition as observed from that direction. The overlap between the two paths is first optimised by coupling the second clock beam into the optical fibre from which the first clock beam is delivered. The probe direction is included in the cycle-by-cycle pseudorandomisation of the servos which acts to suppress any effects from changes in this shift.\par
The measured centre frequency was individually analysed for the two directions in post-processing and no difference was observed within the statistical uncertainty. Nevertheless, an unbounded uncertainty on the shift exists for a non-perfect overlap of the two beams \cite{brewer_27+_2019}. As an alternative means of bounding the shift, we estimate the ion velocity using typical long-term drifts in the level of excess micromotion and imaging of the ion using a CCD camera attached to the optical table. This leads to a drift rate of $\ll 1$~nm/s and a resulting shift of $<1\times10^{-18}$ for each beam, with a negligible uncertainty.\par
Isotropic thermal expansion of the experimental apparatus around the ion could lead to an additional first-order Doppler shift that would not be cancelled by probing with counter-propagating beams. Based on the lab temperature being stabilised to $\pm0.25$~K, we estimate this leads in the worst case to a fractional shift of $5\times10^{-19}$ for a 10~h dataset. This value is taken fully as an additional uncertainty.\par 

\subsection{Second-order Doppler shift}

\subsubsection{Ion temperature}

\begin{table}
	\begin{center}
	    \caption{\textbf{Parameters of the motional modes of \Arthirtysix}}
		\begin{tabular*}{\textwidth}{c @{\extracolsep{\fill}} cccc}
	        \hline
		    \\[-1.3\medskipamount]
			 & $\omega/2\pi$ & $\Bar{n}_0$ & $\Gamma_h$ & \Arthirtysix amplitude\\
			Mode & (MHz) & (quanta) & (quanta/s) & (nm)\\
			\hline
			$x_\mathrm{IP}$  & 5.26 & 0.8(4) & 11.1(10) & 5.2 \\
			$x_\mathrm{OP}$  & 1.29 & 10.8(6) & 1.57(12) & 0.12 \\
			$y_\mathrm{IP}$  & 5.05 & 0.12(6) & 1.65(13) & 5.3 \\
			$y_\mathrm{OP}$  & 1.05 & 21(6) & 0.54(9) & 0.14 \\
			$z_\mathrm{IP}$  & 1.18 & 0.05(1) & 12.4(6) & 5.9 \\
			$z_\mathrm{OP}$  & 1.63 & 0.11(1) & 7.7(5) & 7.8 \\
	        \hline
		\end{tabular*}
		\caption*{The table gives all the motional modes (IP: in-phase, OP: out-of-phase) and the relevant quantities for the induced second-order Doppler shift. The secular frequencies ($\omega$), the mean phonon number ($\Bar{n}_0$) at the beginning of the clock interrogation, the anomalous heating rate ($\Gamma_h$) and the motional amplitude of the ground-state wavefunction of \Arthirtysix are presented. The given errors contain only statistical contributions from quantum projection noise. Measurement procedures are given in the text.}
		\label{tab:motional_shift}
	\end{center}
\end{table}

The nonzero kinetic energy of the HCI after cooling leads to a second-order Doppler shift $\Delta f_\text{QD}$ on the transition frequency relative to the laboratory frame. This shift is given by the sum of the contributions from all of the motional modes $i$:
\begin{equation}
    \frac{\Delta f_\text{QD}}{f_0} = -\frac{1}{2mc^2} \sum_i b_i^2 k_{\text{IMM}, i} \left( \Bar{n}_i + 1/2 \right) h \nu_i,
	\label{eq:secDoppler}
\end{equation}
where $m$ is the ion mass, $c$ is the speed of light in vacuum, $\nu_i = \omega_i / 2\pi$ is the mode frequency, and $\Bar{n}_i$ is the average number of phonons in this mode during the pulse. $b_i$ is the normalised eigenvector component of the HCI. The value of $k_{\text{IMM}, i}$ is 1 and 2 for the axial and radial modes of the ion crystal, respectively, and accounts for the presence of intrinsic micromotion (IMM) in the radial directions in the trap \cite{berkeland_minimization_1998}.\par
The ion is not actively cooled during the clock pulse, leading to a linearly increasing temperature in all modes from anomalous heating. Therefore the values of $\Bar{n}$ for each of the six modes of the two-ion crystal depend on two quantities: the mean number of phonons after ground-state cooling $\Bar{n}_{0}$, and the anomalous heating rate $\Gamma_h$:
\begin{equation}
    \Bar{n}_i = \Bar{n}_{0,i} + \frac{1}{2} \Gamma_{h, i} t_\pi.
\end{equation}
In this work, a probe time of $t_\pi = 15$~ms was used. To determine $\Bar{n}_{0,i}$ and $\Gamma_{h,i}$ we employ sideband thermometry \cite{turchette_heating_2000}. These quantities were measured for all modes before the measurement campaigns, with the results for \Arthirtysix{} presented in Table~\ref{tab:motional_shift} as an example.\par
Deriving the ion temperature from asymmetry in the motional sidebands can lead to incorrect results for nonthermal distributions of motional states that are produced after resolved-sideband cooling \cite{chen_sympathetic_2017, rasmusson_optimized_2021}. Particularly, modes with a large Lamb-Dicke parameter $\eta$ will suffer from population trapping in Fock states where the first-order sideband strength vanishes. For the axial modes, the Lamb-Dicke parameter is relatively large ($\eta_{\mathrm{ax}} \approx 0.5$) which potentially leads to non-negligible population trapping effects, despite us employing second- and third-order sideband cooling pulses \cite{leopold_cryogenic_2019}. For this reason, we performed simulations of our axial sideband cooling sequence using the framework developed in \cite{rasmusson_optimized_2021} to estimate the expected distribution. %
We find that the measured mean phonon number may underestimate the true value by up to a factor of 5. Even in this worst case the shift and its uncertainty remain well below $10^{-18}$ for these modes.\par
The HCI only possesses a small motional amplitude in the radial out-of-phase modes ($b_i^2 \ll 1$). This greatly suppresses the contribution of these modes to its kinetic energy and ground state cooling is not required. We measured a temperature close to the expected Doppler limit for these modes using sideband thermometry on the Raman transition in \Be. To enhance the sensitivity, the 4th-order sidebands are interrogated as they display much greater asymmetry for large $\bar{n}$ \cite{turchette_heating_2000}. The combined fractional shift of both modes is on the order of $10^{-21}$ and therefore negligible.\par
The radial in-phase modes are mostly decoupled from the \Be{} and only the HCI possesses large motional amplitudes in these modes ($b_i^2 \approx 1$) \cite{kozlov_hci}. Resolved-sideband cooling of these modes using the \Be{} is not possible as a consequence of the extremely small Lamb-Dicke parameters. We cool these modes close to their ground states using algorithmic cooling at the beginning of each servo cycle, as described in Ref.~\cite{king_algorithmic_2021}, and then counteract their heating on a pulse-by-pulse basis. At the end of each servo cycle, a single measurement of either the red or blue sideband excitation strength is performed. This allows us to accumulate live data about the temperature and monitor potential issues such as drifts in the mode frequencies. Variation in the ion temperature are taken into account in the uncertainty. The combination of the small Lamb-Dicke parameters for these modes ($\eta_{\mathrm{rad}} \approx 0.05$) and only interleaved algorithmic cooling pulses means that the population distribution is strongly affected by anomalous heating and is therefore expected to be mostly thermal. From the measured values, we estimate the shift from these modes to be $-1.1(4)\times 10^{-18}$ for \Arforty{} and $-1.3(3)\times 10^{-18}$ for \Arthirtysix{}, making it by far the largest shift for any of the secular modes.\par

The heating rate of the axial modes could in principle be eliminated by applying sideband cooling using the \Be{} ion during the clock interrogation. This would not perturb the clock transition in the HCI because of the small differential polarisability. This additional cooling was not implemented for this measurement due to the relatively short clock pulses, which led to a heating of much less than one phonon per mode during this time.\par

\subsubsection{Excess micromotion}

\begin{figure}
	\includegraphics[width=0.8\textwidth]{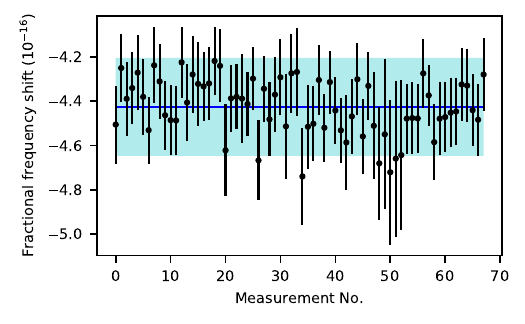}%
	\caption{\textbf{Micromotion-induced frequency shift.} 
	The shown fractional second order Doppler shift derived from micromotion data was taken during the frequency measurement using \Arforty{}, which had a total measurement duration of around 100,000~s split over several days. The blue line indicates the mean value and the shaded area shows the estimated 68\% confidence interval based on the single measurement uncertainty and the standard deviation of the data.}
	\label{fig:mm}
\end{figure}

Oscillating trapping fields at the position of the ion lead to driven motion referred to as excess micromotion. The component of the motion along the line of sight of a particular laser beam can be determined by measurements of the ratio between the Rabi frequencies of the carrier at frequency $\omega_0$ and the first-order micromotion sideband at $\omega_0 + \Omega_{\textrm{RF}}$. Performing this for each of three non-coplanar beams allows the 3D vector of the micromotion to be reconstructed \cite{brewer_27+_2019}. From the measured Rabi frequencies $\Omega_{\textrm{MM}}$ and $\Omega_{\textrm{Car}}$ for the sideband and carrier respectively the modulation index $\beta$ can numerically be calculated using
\begin{equation}
	\frac{\Omega_{\textrm{MM}}}{\Omega_{\textrm{Car}}} = \frac{J_1(\beta)}{J_0(\beta)} = \frac{\beta}{2}+\frac{\beta^3}{16} + \frac{\beta^5}{96} + \frac{17\beta^7}{9216} + ...,
\end{equation}
where $J_i$ are the Bessel functions. For the radial directions we observe $\beta\ll1$ and only the first term in the expansion is required. The measured excess micromotion is dominated by a component along the trap symmetry axis, potentially caused by misalignment of the trap electrodes during its assembly \cite{leopold_cryogenic_2019}. A value of $\beta\approx1.2$ is found in this direction and therefore the higher order terms out to $\beta^7$ are necessary to derive an accurate result.\par
Small changes in radial micromotion were observed on the scale of multiple hours and therefore measurements were made about once per hour during clock operation in order to verify its magnitude. The compensation voltages of the trap were manually re-optimised if a radial displacement of the ion was detected, although changes to the total shift were below the typical uncertainty caused by the axial micromotion alone.\par
This shift is by far the largest in the presented system, and limits the total systematic uncertainty. The measurements of the shift for \Arforty{} are shown in Figure~\ref{fig:mm}. Based on this data the total shift is evaluated to be $-4.43(22)\times10^{-16}$ for this isotope. A slightly larger value of $-5.50(21)\times10^{-16}$ was observed for \Arthirtysix{} under the same trapping conditions, which is consistent with its lower mass compared to \Arforty{}.

\section{Fibre noise}
The clock laser is delivered to the optical table supporting the experimental chamber via a 5~m-long optical fibre. To eliminate effects due to differential motion of the pneumatically floating tables and phase noise introduced by the optical fibre, a small amount of the clock laser light is retroreflected back through the fibre and used for active path-length stabilisation \cite{ma_delivering_1994}. After the retroreflector, around 2~m of free space along with a short 2~m-long fibre delivers the clock laser to the trap itself. These parts of the beam path are passively stabilised using enclosures and the fibre is further insulated using pipe lagging. These measures result in negligible additional noise for the performed experiments. Similarly, the optical fibre that delivers the laser light to the frequency comb laboratory is actively path-length stabilised.\par
To verify that the fibre noise cancellation systems are working correctly, the in-loop beat notes are constantly monitored using frequency counters. To improve the signal-to-noise ratio of the signals to be counted, the signals are amplified and filtered using tracking oscillators. Two trackers are used for each signal in order to detect and remove any cycle slips that may occur. 

\section{Accousto-optical modulator phase chirp}
Pulses from the clock laser beams are produced using dedicated acousto-optical modulators (AOMs). Transient effects after switching on the drive power to the modulator (such as thermal expansion of the AOM crystal) can lead to variations in the laser phase during the pulse that are correlated with the experimental cycle, leading to a systematic offset on the measured transition frequency \cite{degenhardt_influence_2005}. To avoid this, we operate the AOMs at an extremely low drive power of below \SI{5}{\micro\watt}. In Refs.~\cite{rosenband_frequency_2008, kazda_phase_2016} a shift close to or below $10^{-18}$ was observed for several orders of magnitude larger drive power. Therefore this error is negligible at the powers used.

\section{Line pulling}

\subsection{Spectral overlap of different transitions}
The spectrum of the clock transition contains several Zeeman components and their associated motional sidebands, all of which have `wings' that extend from the central resonance. This can lead to an overlap between the various lines, resulting in an asymmetry in the excitation profiles of a particular transition and biasing the servo. Our operational magnetic field was carefully chosen to give a substantial separation between all of the carrier and first-order motional sideband transitions in both the radial and axial directions. As the transition is interrogated with radially-oriented beams, only the radial modes' first-order motional sidebands at ca. $\pm4.5$~MHz are relevant. This is well outside the total Zeeman splitting of around 1~MHz of the carrier transitions and can be safely neglected. The resulting line pulling effect from nearby carrier transitions is $<10^{-22}$ as the $\sim200$~kHz splitting between nearby carrier transitions is large compared to the 50~Hz full width at half maximum of the lines themselves.

\subsection{Clock laser leakage}
\label{subsubsec:clockleakage}
Suppression of the first-order Doppler shift requires interleaved probing from two different directions. Leakage of the incorrect clock laser beam during a pulse could compromise this. To achieve high levels of extinction, the rf signal used to drive the AOM passes through a switch with a nominal isolation of 100~dB. To further reduce the possibility of unwanted diffraction, the AOM drive frequency is detuned far away from resonance after the clock laser pulse has been applied, and only restored shortly before the next pulse begins. An additional problem is that the AOMs that produce the two counterpropagating clock laser beams are operated at nominally identical frequencies. Therefore any pickup of the drive signal for the other AOM could lead to diffraction by the AOM that should be turned off, and cause distortion of the line profile due to unwanted excitation by this laser.\par
To investigate both of these effects, an experimental cycle was run where one of the AOMs was turned on with its drive power at least 15~dB larger than what is used during clock operation, but the beam was physically blocked. The other AOM was left turned off, but with the beam path to the ion free. In this worst-case scenario, any excitation of the clock transition during the cycle would indicate unwanted diffraction by the second AOM. No excitation of the clock transition was observed for interrogation periods of up to 200~ms, far exceeding the typical probe time of 15~ms. 

\subsection{Logic laser leakage}
\label{subsubsec:logicleakage}
Similarly to potential leakage of the `wrong' clock laser beam as described in the previous subsection, any leakage of one of the logic laser beams could lead to unwanted excitation during the clock pulse or during the driving of the much weak micromotion sidebands. To improve the extinction during the clock pulse, both a single-pass and double-pass AOM are switched off, and their drive frequencies are greatly detuned in a similar manner to the clock laser. During the logic pulses, only the single-pass AOM is used for extinction as the double-pass AOM is common to all beams. A similar test was performed as detailed for the clock laser, where a normal experimental cycle was run but the clock beams were physically blocked. No excitation of the clock transition was observed for interrogation periods of up to 200~ms.  

\section{Servo error}
\label{subsec:servoerr}
The second-order integrating servo allows compensation for linear drifts between the atomic transition frequency and the laser frequency \cite{peik_laser_2006}. Higher-order drifts can still lead to the locked laser frequency being offset from the centre of the atomic transition. As the linear drift of the cryogenic silicon cavity leads to a drift of only a few tens of hertz on the clock laser frequency over the course of the day, with higher-order drifts being even lower in amplitude, the dominant source of instability on relevant timescales are therefore changes in the magnetic field.\par
As discussed earlier, we observe a field stability of around 0.2~nT at short timescales at the position of the ion, which corresponds to a frequency instability of 5~Hz for the most sensitive lines, with variations of a few nT over the course of a day. If the Zeeman components were always interrogated in the same order, a constant drift of the magnetic field between interrogation of the opposing Zeeman components would lead to an incomplete cancellation of the shift when the two frequencies are averaged. This effect is mitigated by pseudo-randomising the order in which the Zeeman components are interrogated on a cycle-by-cycle basis.\par
Any residual offset of the servos from the centres of their respective lines was investigated in two ways. Firstly, it was confirmed that the mean error signals for each of the individual servos during the measurement was consistent with zero. Secondly, simplified simulations of the servo behaviour in the presence of an unstable magnetic field were performed. In order to see the response of the system to nonlinear, asymmetric drifts that do not average to zero over long measurements, a combination of linear drift at a rate of 2~pT/s and a quadratic drift of 0.015~fT/s$^2$ was applied between interrogations. Both types of drift would result in a long-term instability ($>$1000~s) that is at least an order of magnitude higher than we typically observe in our apparatus. 1000 repetitions of the simulation, each with an equivalent lock time of 85,000~s, indicate offsets of approximately $\pm1$~mHz for the servos tracking the most sensitive Zeeman components. These offsets are equal and opposite for pairs of Zeeman components, with the resulting effect on the mean of the eight servos being $0 \pm 0.2$~mHz. This shift is therefore negligible under the much smaller magnetic field changes observed in the experiment.

\section{Background gas collisions}
A cryogenic environment is used to achieve a vacuum level that is orders-of-magnitudes lower than room-temperature systems. The collision rate scales linearly with the background gas particle density which suppresses a potential frequency shift due to collisions by orders of magnitude compared to room-temperature systems. Additionally, in contrast to frequency standards based on neutral and singly-charged atoms, collisions between background gas particles and an HCI are catastrophic and lead to a change of its charge state and are equivalent to a loss of the ion from the trap. %
All of these factors render this shift negligible.

\section{Effect of frequency comparison process}
The frequency combs themselves introduce a systematic shift of less than $5\times10^{-21}$ \cite{benkler_end--end_2019}. However, inside the frequency comb laboratory there are a few metres of free-space beam path and optical fibres that are not actively path-length stabilised. This leads to a typical excess instability of $1\times10^{-18}$ at our measurement duration of 100,000~s. We take this value as an additional statistical uncertainty as a conservative estimate. The computation of the optical frequency ratio from the various counted beat frequencies has been carried out using arbitrary precision number formats and alternatively following the approach in Ref.~\cite{lodewyck_universal_2020} to ensure that the result is not limited by the numerical resolution.

\section{Gravitational redshift}
The \Yb{} and \Ar{} ions are located in separate buildings, and at different elevations above the geoid. This height difference leads to a gravitational redshift that must be taken into account. As part of the EMRP project SIB55 `International timescales with optical clocks', local surverying measurements were carried out to determine the height difference between the the \Yb{} laboratory and a laboratory holding a $^{27}$Al$^{+}$ atomic clock, which is located four floors higher in the same building as the HCI experiment. The difference in height between the HCI and the $^{27}$Al$^{+}$ reference point was determined to be 11.13(5)~m using a laser range finder and a laser level. The uncertainty accumulates from the device accuracy as well as angle uncertainties when using the laser range finder. The height difference leads to a fractional shift of $1.64(6)\times 10^{-16}$ for the \Ar{} frequency relative to the \Yb{} reference frame. 

\section{Isotope shift calculations}

\subsection{Relativistic mass shift}

The calculation of the mass shift of the $^2$P$_{1/2}$--$^2$P$_{3/2}$ transition of B-like \Ar{} was carried out by the configuration-interaction (CI) method, as implemented in our previous works (see Ref.~\cite{yerokhin_nonlinear_2020} and references therein). This is complicated by a strong mixing of the reference $1s^22s^22p_j$ state with the closely-lying $1s^22p^3$ excited states. In order to take this into account, we constructed the CI expansion from multiple reference states, specifically $1s^22s^22p_{1/2} + 1s^22p_{1/2}2p_{3/2}^2$ for $J = 1/2$ and $1s^22s^22p_{3/2} + 1s^22p_{1/2}^22p_{3/2} + 1s^22p_{3/2}^3 $ for $J = 3/2$. The CI expansions included single, double, triple, and quadruple excitations from the multiple reference states specified above. It is noteworthy that double excitations from a multiple reference state are equivalent to a subset of quadruple excitations from a single reference state.

Results of our numerical calculations of the relativistic mass-shift constant are presented in Table~\ref{tab:CI}. We present them for two different choices of one-electron orbitals (labelled as Basis \#1 and Basis \#2 in the table). For Basis \#1, we used one-electron orbitals from the ($B$-spline finite-basis representation of the) Dirac spectrum obtained with the frozen-core Dirac-Fock (FCDF) potential. For Basis \#2, we use the exact Dirac-Fock orbitals for the $n = 1$ and $n = 2$ states and the orthonormalized FCDF orbitals for the virtual excited states. For comparison, the third column of Table~\ref{tab:CI} contains results obtained with the single reference state and Basis \#2.

We observe that when one constructs the CI expansion with a single reference state, the contribution of triple and quadruple excitations becomes orders of magnitude larger than for a multiple reference state. The final CI result for the single reference state is consistent with that obtained with the multiple reference state, but much less accurate. It also is remarkable that, if we restrict the space of the configuration-state wave functions to the single and double excitations only, we arrive at different results for Basis \#1 and Basis \#2. Only after triple excitations are properly accounted for, the agreement between different choices of the basis is restored.

Another remarkable feature of our results is a large contribution of the quadruple excitations. As seen from Table~\ref{tab:CI}, it turns out to be typically larger than that of the triple excitations, and of opposite sign. Because of this large cancellation, it was important to use the same sets of one-electron orbitals in computations of contributions induced by the triple and quadruple excitations.

\begin{table}[t]
    \caption{\textbf{Numerical results of the CI calculation of the relativistic mass shift constant $K$ for the $^2$P$_{1/2}$--$^2$P$_{3/2}$ transition in B-like \Ar{}, in a.u.}}
    \begin{tabular*}{\textwidth}{l @{\extracolsep{\fill}}w{1.7}w{1.7}w{1.6}}
        \hline
        \multicolumn{1}{l}{Excitations} & \multicolumn{2}{c}{Multiple reference state} & \multicolumn{1}{c}{Single} \\
                                        & \multicolumn{1}{c}{Basis \#1}
                                                    & \multicolumn{1}{c}{Basis \#2} & \multicolumn{1}{c}{reference state} \\
        \hline
        Single \& Double & -0.189\,90       &  -0.189\,61        &  -0.188\,0       \\
        Triple           &  0.000\,40       &   0.000\,18        &   0.002\,4       \\
        Quadruple        & -0.000\,44       &  -0.000\,43        &  -0.012\,9       \\
        Sum              & -0.189\,94\,(4)  &  -0.189\,87\,(13)  &  -0.189\,6\,(8)  \\
        Final            & -0.189\,94\,(8) \\
        \hline
    \end{tabular*}
    \caption*{The conversion factor from a.u. to units of 1000$\times$GHz$\times$amu 
    used in Ref.~{\cite{zubova_isotope_2016}} is $3.609\,482$.
    \label{tab:CI}}
\end{table}

\subsection{QED mass shift}

The QED theory of the nuclear recoil effect (to first order in $m/M$ but to all order in $\Za$) was developed by V.~M.~Shabaev in Refs.~\cite{shabaev_mass_1985,shabaev_qed_1998}. Numerical calculations of the QED nuclear recoil for few-electron atoms were performed to the leading order in $1/Z$, $1/Z^0$, in Refs.~\cite{artemyev_relativistic_1995,artemyev_nuclear_1995,malyshev_nuclear_2018} and to the first order in $1/Z$ in Ref.~\cite{malyshev_qed_2019}.

In analogy to the relativistic mass shift, the QED correction to the mass shift constant $K$ can be conveniently separated into the normal mass shift (NMS) and the specific mass shift (SMS) parts,
\begin{align}
K_{\rm rec, qed} = K_{\rm nms, qed} + K_{\rm sms, qed}\,.
\end{align}
To leading order in the parameter $1/Z$, the theoretical treatment is reduced to  the independent-electron approximation, in which the normal QED mass shift is just a sum of one-electron contributions,
\begin{align}
K_{\rm nms, qed} = \sum_i \frac{(\Za)^5}{\pi n_i^3}\,P_i(\Za)\,,
\end{align}
with the sum over $i$ running over all one-electron orbitals and $n$ being the principal quantum number of the one-electron state. Explicit formulae and the method of calculation of the function $P(\Za)$ are described in the literature \cite{artemyev_relativistic_1995,yerokhin_nuclear_2015,yerokhin_nuclear_2016}.

The specific QED mass shift in the independent-electron approximation for an alkali-like atomic state is given by \cite{artemyev_relativistic_1995}
\begin{align}
K_{\rm sms, qed} &\ = \sum_c \Big\{ \big<v\vert\bfp\,\vert c\big>\big<c\vert\big[\bm{D}(\Delta)-\bm{D}(0)\big]\vert v\big>
                + \big<v\vert\big[\bm{D}(\Delta)-\bm{D}(0)\big]\vert c\big>\big<c\vert\bfp\,\vert v\big>\Big\} \nonumber \\
     &\ - \sum_c \big<v\vert\bm{D}(\Delta)\vert c\big>\,\big<c\vert\bm{D}(\Delta)\vert v\big>\,,
\end{align} where the summation over $c$ runs over the core electron states, $v$ denotes the valence electron state, $\Delta = \vare_v-\vare_c$,
\begin{align}
D_k(\omega,\bfr) = \Za\,\alpha_l\bigg[ \frac{e^{i\vert\omega\vert r}}{r}\,\delta_{lk} + \frac{\partial}{\partial{r_l}}
\frac{\partial}{\partial{r_k}}
\frac{e^{i\vert\omega\vert r}-1}{\omega^2 \, r}\bigg]\,,
\end{align}
$\balpha$ is the vector of Dirac matrices, and $r = \vert\bfr\vert$. In the limit $\omega \to 0$, $D(\omega)$ takes the form
\begin{align}
D_k(0,\bfr) = \frac{\Za}{2r}\bigg[\alpha_k + \frac{(\balpha\cdot\bfr)\, r_k}{r^2} \bigg]\,.
\end{align}

In the present work, we calculate the QED recoil effect within the independent-electron approximation, with one-electron wave functions calculated in the localised Dirac-Fock (LDF) potential. Previously, the same approach was used by Zubova et al.~\cite{zubova_isotope_2016}. The numerical method follows the approach of Refs.~\cite{yerokhin_nuclear_2015,yerokhin_nuclear_2016}, and the construction of the LDF potential is performed as described in Ref.~\cite{yerokhin_qed_2007}. Our numerical result is presented in Table~4 of the main text, in agreement with Ref.~\cite{zubova_isotope_2016}. The uncertainty of 20\% ascribed to the QED recoil correction represents the influence of residual electron-correlation effects; it was obtained by comparing the result for the relativistic recoil effect delivered by the LDF potential with the full CI result.

\subsection{Radiative mass shift}

The contribution of the pure radiative recoil effect (of order $\alpha\, m/M$) to the mass shift is very small. Indeed, the estimation based on the hydrogen theory gives a result proportional to the field-shift constant $F$ (see Eq.~(47) of Ref.~\cite{yerokhin_nuclear_2015}), $\delta K \sim \alpha\, (\Za)\,F$, which is negligible. There is, however, a much larger contribution coming from QED shifts of the energy levels. This effect is enhanced due to strong mixing of the reference state with the closely-lying virtual excited states. The QED effects alter positions of energy levels, and thus influence the level mixing which is proportional to the inverse energy intervals.

We calculated the correction to the mass-shift constant due to QED energy shifts by including the effective QED operator \cite{shabaev_model_2013,shabaev_qedmod_2015} into the Hamiltonian matrix of the CI method, and repeating our calculations with and without the QED addition. The numerical result is included into the entry `Mass shift, QED' of Table 4 of the main text.

\subsection{QED field shift}

In the present work we accounted for the leading QED effects to the field-shift constant. It was found that the leading QED correction comes from QED shifts of energy levels. This effect is enhanced due to strong mixing of the reference states with the closely-lying virtual excited states. We calculated the QED correction due to energy shifts by including the effective QED operator \cite{shabaev_model_2013,shabaev_qedmod_2015} into the Hamiltonian matrix of the CI method and repeating our calculations with and without the QED addition. 

The second-largest QED effect is the pure radiative finite nuclear size (fns) correction. Following Ref.~\cite{zubova_isotope_2016}, we estimate it by rescaling the known hydrogenic radiative fns result for the $1s$ state, as follows:
\begin{align}
F_{\rm rad} \approx -0.9 \, \frac{\alpha}{\pi}\,F_{\rm rel}\,.
\end{align}
The numerical coefficient in the above equation is the radiative fns correction for $Z = 18$ and the $1s$ state calculated in Ref.~\cite{yerokhin_nuclear-size_2011} (specifically, $G_{\rm NSE}+G_{\rm NVP}$, see Eqs.~9 and 22 of that work). It is remarkable that the QED contribution from energy shifts is significantly larger than what could be expected from the hydrogenic approximation. Strictly speaking, there is some double counting between the two QED corrections considered above. However, its effect is negligible at the level of accuracy we are presently concerned with. The numerical result for the QED field shift is presented in Table 4 of the main manuscript.